\documentclass[prd,aps,showpacs,nofootinbib,preprint,eqsecnum]{revtex4}
\usepackage{amsmath}
\usepackage{amssymb}
\usepackage{amsfonts}
\usepackage{graphicx,bm}
\usepackage{dcolumn}
\usepackage{color,amsxtra}
\usepackage{epsf}
\usepackage{enumerate}
\usepackage{hhline}
\usepackage{array}
\usepackage{tabularx}
\usepackage{subfigure}
\usepackage{fancyhdr}
\usepackage{mathrsfs}
\newcommand{\be}{\begin{equation}}
\newcommand{\ee}{\end{equation}}
\newcommand{\bea}{\begin{eqnarray}}
\newcommand{\eea}{\end{eqnarray}}
\newcommand{\beaa}{\begin{eqnarray*}}
\newcommand{\eeaa}{\end{eqnarray*}}

\newcommand{\Eqn}[1]{&\hspace{-0.2em}#1\hspace{-0.2em}&}

\def\be{\begin{equation}}
\def\ee{\end{equation}}
\def\bea{\begin{eqnarray}}
\def\eea{\end{eqnarray}}

\begin{document}
%\draft

%\tolerance=5000

%%%%%%%%%%%%%%%%%%%%%
%  Title
%%%%%%%%%%%%%%%%%%%%%
\title{Growth of matter perturbations for realistic $F(R)$ models}

\author{Antonio Jes\'us L\'opez-Revelles\footnote{alopez@ieec.uab.es}}
\affiliation{Consejo Superior de
Investigaciones Cient\'{\i}ficas, ICE/CSIC-IEEC, Campus UAB,
Facultat de Ci\`{e}ncies, Torre C5-Parell-2a pl, E-08193 Bellaterra
(Barcelona), Spain}

%\date{}

%\def\thesection{\Roman{section}}
%\def\theequation{\Roman{section}.\arabic{equation}}

%%%%%%%%%%%%%%%%%%%%%
% Abstract
%%%%%%%%%%%%%%%%%%%%%
\begin{abstract}

    Two different realistic $F(R)$ modified gravity models are considered in the framework of the Friedmann-Lemetre-Robertson-Walker
    universe. The parameters of these two models are adjusted to reach coherence with the most recent and accurate observations of the current universe.
    A study of the growth of matter density perturbations is done, and several parametrizations of the growth index are developed
    for both models. The ansatz for the growth index given by $\gamma = \gamma_0 + \gamma_1 z/(1+z)$ seems to be the best
    parametrization for the two models considered. Finally, the values obtained for $\gamma_0$ and $\gamma_1$ can be used in
    order to characterize these two models and to differentiate them from others such as the Hu-Sawicki model.

\end{abstract}
%%%%%%%%%%%%%%%%%%%%%

%----------------------------
%PACS
%----------------------------

\maketitle
%===========================================================================

%%%%%%%%%%%%%%%%%%%%%%%%%%%
%%%  Sec. I
%%%%%%%%%%%%%%%%%%%%%%%%%%%
\section{Introduction \label{SectI}}

    Over the last years, several observations such as supernovae Ia (SNe Ia)~\cite{SN1}, large scale structure (LSS)~\cite{LSS} with
    baryon acoustic oscillations (BAO)~\cite{Eisenstein:2005su}, cosmic microwave background (CMB) radiation
    ~\cite{WMAP-Spergel, WMAP, Komatsu:2010fb} and weak lensing~\cite{Jain:2003tba} have demonstrated that our Universe is suffering from an
    acceleration in its expansion. The explanation for this accelerated expansion has become one of the most important theoretical
    problems for the scientific community. Over the last years, several suggestions have been made in attempt to solve this problem.
    One of these suggestions introduces an exotic kind of energy, called dark energy, into the framework of the theory of
    general relativity (for recent reviews in terms of dark energy, see~\cite{Li:2011sd, Kunz:2012aw, Bamba:2012cp}).
    There is another group of theories that try to give an explanation to this late-time cosmic acceleration, which is based on the modification of
    Einstein's gravity (for recent reviews on modified gravity, see~\cite{Review-Nojiri-Odintsov, Viableconditions, Clifton:2011jh,
    Capozziello:2011et, Capozziello:2012hm}). $F(R)$ modified gravity is among these last kinds of theories.

    The natural unification of inflation and late-time cosmic acceleration achieved by some $F(R)$ modified gravity theories make them
    very attractive (see \cite{Nojiri:2007}). It has also been demonstrated that some of these theories can be suitable
    candidates to explain the Universe we live in. In this sense, some of these $F(R)$ gravity theories can pass the Solar System tests
    (see \cite{Nojiri:2003ft,Chiba:2003ir,SolarSystemconstraints}), and they can reproduce the $\Lambda$-cold-dark-matter ($\Lambda$CDM)
    universe in the large curvature regime. These theories can also accomplish the matter stability condition (see
    \cite{Nojiri:2007, Faraoni, Song:2006ej}), and they can have a stable late-time de Sitter point (see \cite{D-S-P,Cognola:2008zp}).
    Different observational manifestations of $F(R)$ gravity were studied in \cite{listone}.

    As a consequence of the large number of different gravitational theories, a problem of distinction among some of them has appeared. The fact that
    different models can achieve the same expansion history has revealed that another tool, which may provide a way for
    discriminating among different gravitational theories, may be required. The study of the growth of matter density perturbations may become the
    tool that we need, due to the fact that theories with the same expansion history can have a different cosmic growth history. In order to
    characterize the growth of matter density perturbations, the so-called growth index $\gamma$ (see \cite{Linder:2005in}) can be
    very useful.

    In this work, the study of the growth history has been done for two different $F(R)$ modified gravity models, and the growth index has been determined for both
    models. The paper is organized as follows. In Sec.~II, the formulations and the dynamics of modified $F(R)$ gravity in the framework
    of a Friedmann-Lemetre-Robertson-Walker (FLRW) universe is briefly reviewed~\cite{cno06}. In Sec.~III, two different modified $F(R)$
    gravity models are considered, and the values of the parameters are adjusted for both models to reach coherence with
    recent observations of the Universe. In Sec.~IV, the study of the growth of matter density perturbations is done for these
    two $F(R)$ gravity models, and several parametrizations for the growth index are studied for both models. Finally, a summary for this work is given in Sec.~V.

%%%%%%%%%%%%%%%%%%%%%%%%%%%
%%%  Sec. II
%%%%%%%%%%%%%%%%%%%%%%%%%%%
\section{Dynamics of $F(R)$ gravity in the FLRW universe}

    In this section, the formulation of modified $F(R)$ gravity is reviewed in the framework of the FLRW universe (see \cite{Bamba:2012qi}). The general action
    considered for modified $F(R)$ gravity is given by
        \begin{equation}\label{ii1}
            I = \int \limits_{\mathcal{M}} d^4x \, \sqrt{-g} \, \left( \frac{F(R)}{2 \kappa^2} + \mathcal{L}_{matter} \right),
        \end{equation}
    where $\mathcal{M}$ denotes the space-time manifold, $g$ denotes the determinant of the metric tensor $g_{\mu \nu}$, $F(R)$ is a
    generic function of the Ricci scalar $R$, $\mathcal{L}_{matter}$ denotes the matter Lagrangian and $\kappa^2 = 8 \pi G_N \equiv 8
    \pi / M^2_{Pl}$ is the gravitational constant, with the Planck mass being $M_{Pl} = G^{-1/2}_N = 1.2 \times 10^{19} GeV$. Note
    that, throughout this paper, the use of natural units will be assumed. Therefore, $\kappa_B = c = \hbar = 1$. Finally, the generic
    function $F(R)$ will be given by
        \begin{equation}\label{ii2}
            F(R) = R + f(R),
        \end{equation}
    with $f(R)$ another generic function of $R$. It might be accurate here to point out that the $R$ term in the expression of $F(R)$ accounts for the
    Einstein-Hilbert action in the theory of general relativity, while the function $f(R)$ encodes the modification of gravity.

    The variation of the action (\ref{ii1}) with respect to the metric $g_{\mu \nu}$ gives the field equations for modified $F(R)$
    gravity
        \begin{equation}\label{ii3}
            R_{\mu \nu} - \frac{1}{2} R g_{\mu \nu} = \frac{1}{F'(R)} \left[ \frac{1}{2} g_{\mu \nu} \left( F(R) - R F'(R) \right) +
            \left( \nabla_\mu \nabla_\nu - g_{\mu \nu} \Box \right) F'(R) \right] + \frac{\kappa^2}{F'(R)} T_{\mu \nu}^{matter},
        \end{equation}
    where $R_{\mu \nu}$ is the Ricci tensor, the prime denotes the derivative with respect to the Ricci scalar $R$, $\nabla_\mu$ is
    the covariant derivative operator associated with the metric $g_{\mu \nu}$, $\Box \phi = g^{\mu \nu} \nabla_\mu \nabla_\nu \phi$
    is the covariant d'Alembertian operator for a scalar field and $T^{matter}_{\mu \nu} = \mbox{diag}(\rho_m,P_m,P_m,P_m)$ is the
    matter stress-energy tensor, with $\rho_m$ and $P_m$ being the energy density and pressure of matter, respectively.

    The trace of Eq.(\ref{ii3}) yields
        \begin{equation}\label{ii4}
            \Box F'(R) = \frac{\partial V_{eff}}{\partial F'(R)},
        \end{equation}
    where
        \begin{equation}\label{ii5}
            \frac{\partial V_{eff}}{\partial F'(R)} = \frac{1}{3} \left( 2 F(R) - R F'(R) + \kappa^2 T^{matter} \right),
        \end{equation}
    with $T^{matter} = g^{\mu \nu} T^{matter}_{\mu \nu}$ being the trace of the matter stress-energy tensor and $F'(R)$ being the
    so-called ``scalaron'', i.e., effective scalar degree of freedom.

    From now on, flat FLRW space-time is assumed, with the metric given by
        \begin{equation}\label{ii6}
            ds^2 = - N(t)^2 dt^2 + a(t)^2 d{\bf x}^2,
        \end{equation}
    where $N(t)$ is a function of the cosmic time $t$, which will be taken as $N(t) = 1$ in the following, and $a(t)$ is the so-called
    scale factor. The Ricci scalar in this metric can be written as
        \begin{equation}\label{ii7}
            R = 12 H^2 + 6 \dot H,
        \end{equation}
    where $H$ is the Hubble parameter, which is related to the scale factor $a(t)$ through the expression $H = \dot a(t) / a(t)$, and
    the dot denotes the derivative with respect to the cosmic time $t$.

    From (\ref{ii3}) two gravitational field equations can be obtained by considering the component $(\mu,\nu) = (0,0)$ and the trace
    of $(\mu,\nu)=(i,j)$ (with $i,j=1,2,3$). These equations are given by
        \begin{eqnarray}\label{ii8}
            \frac{3 H^2}{\kappa^2} = \frac{1}{F'(R)} \left\{ \rho_m + \frac{1}{2 \kappa^2} \left[ R F'(R) - F(R) - 6 H \dot F'(R)
            \right] \right\},
            \\
            - \frac{2 \dot H + 3 H^2}{\kappa^2} = \frac{1}{F'(R)} \left\{ p_m + \frac{1}{2 \kappa^2} \left[ - R F'(R) + F(R) + 4 H
            \dot F'(R) + 2 \ddot F'(R) \right] \right\}.
        \end{eqnarray}
    By introducing the effective energy density, $\rho_{eff}$, and the effective pressure, $P_{eff}$, as
        \begin{eqnarray}\label{ii9}
            \rho_{eff} \equiv \frac{1}{F'(R)} \left\{ \rho_m + \frac{1}{2 \kappa^2} \left[ R F'(R) - F(R) - 6 H \dot F'(R)
            \right] \right\},
            \\
            P_{eff} \equiv \frac{1}{F'(R)} \left\{ p_m + \frac{1}{2 \kappa^2} \left[ - R F'(R) + F(R) + 4 H
            \dot F'(R) + 2 \ddot F'(R) \right] \right\},
        \end{eqnarray}
    Eq.(\ref{ii8}) can be written as
        \begin{eqnarray}\label{ii10}
            \frac{3 H^2}{\kappa^2} = \rho_{eff},
            \\
            - \frac{2 \dot H + 3 H^2}{\kappa^2} = P_{eff},
        \end{eqnarray}
    which are the same equations that can be obtained in the framework of the theory of general relativity by just changing the energy density
    and the pressure of matter by these new effective energy density and effective pressure.

    In order to study the dynamics of modified $F(R)$ gravity models in the framework of a flat FLRW universe, it may be useful to introduce
    the new function $y_H(z)$ given by
        \begin{equation}\label{ii11}
            y_H(z) \equiv \frac{\rho_{DE}}{\rho_{m(0)}} = \frac{H^2}{\tilde{m}^2} - (1 + z)^3 - \chi (1 + z)^4,
        \end{equation}
    with $z = 1/a(t) - 1$ being the redshift, $\rho_{m(0)}$ being the matter energy density at the present time, $\tilde{m}^2$ being
    the mass scale given by
        \begin{equation}\label{ii12}
            \tilde{m}^2 \equiv \frac{\kappa^2 \rho_{m(0)}}{3} \simeq 1.5 \times 10^{-67} eV^2
        \end{equation}
    and, finally, $\chi$ being defined as
        \begin{equation}\label{ii13}
            \chi \equiv \frac{\rho_{r(0)}}{\rho_{m(0)}} \simeq 3.1 \times 10^{-4},
        \end{equation}
    where $\rho_{r(0)}$ is the radiation energy density at the present cosmic time.

    Introducing Eq.(\ref{ii11}) into the first equation in (\ref{ii10}) yields
        \begin{equation}\label{ii14}
            \frac{d^2 y_H(z)}{d z^2} + J_1 \frac{d y_H(z)}{d z} + J_2 \left( y_H(z) \right) + J_3 = 0,
        \end{equation}
    where
        \begin{eqnarray}
            J_1 \Eqn{=} \frac{1}{(z+1)} \left[ - 3 - \frac{1}{y_H + (z + 1)^{3} + \chi (z + 1)^{4}} \frac{1 - F'(R)}{6 \tilde{m}^2
            F''(R)} \right],
            \\
            J_2 \Eqn{=} \frac{1}{(z + 1)^2}\left[ \frac{1}{y_H + (z + 1)^{3} + \chi (z + 1)^{4}} \frac{2 - F'(R)}{3 \tilde{m}^2
            F''(R)} \right],
            \\
            J_3 \Eqn{=} - 3 (z + 1)
            \nonumber \\
            && - \frac{(1 - F'(R))((z + 1)^{3} + 2 \chi (z + 1)^{4}) + (R - F(R))/(3 \tilde{m}^2)}{(z + 1)^2 (y_H + (z + 1)^{3} + \chi
            (z + 1)^{4})} \frac{1}{6 \tilde{m}^2 F''(R)}.
        \end{eqnarray}

    Furthermore, the Ricci scalar is expressed as
        \begin{equation}\label{ii15}
            R = 3 \tilde{m}^2 \left[ 4 y_H(z) - (z + 1) \frac{d y_H(z)}{dz} + (z + 1)^{3} \right].
        \end{equation}

    In deriving this equation, we have used the fact that $- (z + 1) H(z) d/dz = H(t) d/d(\ln a(t)) = d/dt$, where $H$ could be an
    explicit function of the redshift as $H = H(z)$, or an explicit function of the time as $H = H(t)$.

    Once the function $y_H(z)$ is obtained from Eq.(\ref{ii14}), it can be used to calculate the equation of state of dark
    energy $\omega_{DE}$ through the expression given by
        \begin{equation}\label{ii16}
            \omega_{DE}(z) \equiv \frac{P_{\mathrm{DE}}}{\rho_{\mathrm{DE}}} = - 1 + \frac{1}{3} (z+1) \frac{1}{y_H(z)}
            \frac{d y_H(z)}{d (z)},
        \end{equation}
    and the dark energy density parameter $\Omega_{DE}$ given by
        \begin{equation}\label{ii17}
            \Omega_{DE}(z) \equiv \frac{\rho_\mathrm{DE}}{\rho_\mathrm{eff}} =
            \frac{y_H}{y_H+\left(z+1\right)^3+\chi\left(z+1\right)^4},
        \end{equation}
    which are two essential quantities in determining whether a theory is realistic or not.

%%%%%%%%%%%%%%%%%%%%%%%%%%%
%%%  Sec. III
%%%%%%%%%%%%%%%%%%%%%%%%%%%
\section{Realistic $F(R)$ models}

    In this section, two different kinds of modified $F(R)$ gravity models will be considered. The parameters of these models will be set in order to reproduce
    recent observations of our current Universe.

    In \cite{Komatsu:2010fb}, Komatsu {\it et al.} determined important cosmological parameters by combining the seven-year WMAP data with the latest distance
    measurements from the (BAO) in the distribution of galaxies, the Hubble constant ($\mathrm{H}_0$) measurement and the last
    observations coming from the luminosity distances out to high-z type Ia supernovae (SN). The determined values for the dark energy equation of state parameter
    $\omega_{DE}$ and for the dark energy density parameter $\Omega_{DE}$ are given by
        \begin{equation}\label{iii6}
            \omega_{DE} = -0.980 \pm 0.053 \ \mbox{from (WMAP+BAO+SN)} \ ,$$
            $$\Omega_{DE} = 0.725 \pm 0.016 \ \mbox{from (WMAP+BAO+}\mathrm{H}_0) \ .
        \end{equation}
    From now on, these results will be used as a constraint for the two model parameters.

    \subsection{First $F(R)$ model}

    In the first place, a model that appeared in \cite{Nojiri:2007cq} which could unify inflation and current acceleration will be considered.
    This model is given by
        \begin{equation}\label{iii1}
            F(R) = R + \frac{\alpha R^{m + l} - \beta R^n}{1 + \gamma R^l},
        \end{equation}
    where $\alpha$, $\beta$ and $\gamma$ are positive constants and $m$, $n$ and $l$ are positive integers satisfying the condition
    $m + l > n$. The model given by (\ref{iii1}) is a generalization of the Hu-Sawicki model (see \cite{Hu:2007nk}) which has been
    proposed by Hu and Sawicki as a model which is in agreement with the constraints imposed by the Solar System tests. Model
    (\ref{iii1}) can be reparametrized by choosing $n = l$, $\beta = 2 \Lambda/(b \Lambda)^n$ and $\gamma = 1/(b \Lambda)^n$ yielding
        \begin{equation}\label{iii2}
            F(R) = R - 2 \Lambda \left( 1 - \frac{1}{1 + \left( \frac{R}{b \Lambda} \right)^n} \right) +
            \frac{\alpha R^{m + n}}{1 + \left( \frac{R}{b \Lambda} \right)^n},
        \end{equation}
    with $\Lambda$ being the current cosmological constant. It is worth noting that the new constant $b$ can be negative
    when $n$ is a positive even; in any other it must be positive.

    The next step should be to solve Eq.(\ref{ii14}) for model (\ref{iii2}) and to find out the constraints on the values of the model parameters needed
    to fulfill the conditions given by (\ref{iii6}). Unfortunately, because Eq.(\ref{ii14}) cannot be solved in an analytical way for the $F(R)$ model given by
    (\ref{iii2}), this is not possible. Thus, the way to solve this problem is to suggest a set of parameters for (\ref{iii2}), to solve Eq.(\ref{ii14})
    numerically and to check if the results are in accordance with (\ref{iii6}).

    For the model given by (\ref{iii2}), the following values for the parameters have been chosen:
        \begin{equation}\label{iii7}
            n = 4, \quad m = 1, \quad b = \frac{3}{8}, \quad \alpha = 10^{-10} \tilde{m}^{-8},
        \end{equation}
    with $\Lambda = 7.93 \tilde{m}^2$ in accordance with \cite{Komatsu:2010fb}. In order to obtain the initial conditions needed to
    solve Eq.(\ref{ii14}) numerically, we may evaluate the dark energy density $\rho_{DE} = \rho_{eff} - \rho_m$ from Eq.(\ref{ii9})
    at the matter dominated era (high redshifts) by putting $R = 3 \tilde{m}^2 (1 + z)^3$. In the case of the first model with the set
    of parameters given by (\ref{iii7}), the initial conditions can be written as
        \begin{equation}\label{iii8}
            \left. y_H(z) \right|_{z_i} = \frac{\Lambda}{3 \tilde{m}^2} - 81 \alpha \tilde{m}^4 (1 + z)^3,$$
            $$\left. \frac{dy_H(z)}{dz} \right|_{z_i} = - 243 \alpha \tilde{m}^4 (1 + z)^2.
        \end{equation}
    In the case of model (\ref{iii2}) with the set of parameters given by (\ref{iii7}) and the initial conditions given by
    (\ref{iii8}), I set $z_i = 3.40$, obtaining $\omega_{DE}(0) = -1.000$ and $\Omega_{DE} = 0.725$, which are in accordance
    with the observational data given by (\ref{iii6}). Note that it is hard to solve Eq.(\ref{ii14}) for higher values of the
    redshift due to the large frequency of the dark energy oscillations.

    In the following, model (\ref{ii2}) with the set of parameters given by (\ref{iii4}) will be called {\bf model I}.

    \subsection{Second $F(R)$ model}

    The second model considered \cite{Nojiri:2010ny} is given by
        \begin{equation}\label{iii3}
            F(R) = R - \alpha_0 \left[ \tanh{\left( \frac{b_0 (R - R_0)}{2} \right)} + \tanh{\left( \frac{b_0 R_0}{2} \right)}
                \right] - \alpha_I \left[ \tanh{\left( \frac{b_I (R - R_I)}{2} \right)} + \tanh{\left( \frac{b_I R_I}{2} \right)}
                \right].
        \end{equation}
    It will be assumed that $R_I \gg R_0$, $\alpha_I \gg \alpha_0$ and $b_I \ll b_0$, with $b_I R_I \gg 1$. By choosing $2 \Lambda_0 = \alpha_0 \left[ 1 +
    \tanh{\left( \frac{b_0 R_0}{2} \right)} \right]$ and $ 2 \Lambda_I = \alpha_I \left[ 1 + \tanh{\left( \frac{b_I R_I}{2} \right)} \right]$, Eq.(\ref{iii3})
    reduces to
        \begin{equation}\label{iii4}
            F(R) = R - 2 \Lambda \left[ 1 - \frac{1 - \tanh{\left( \frac{b_0 (R - R_0)}{2} \right)}}{1 + \tanh{\left(
            \frac{b_0 R_0}{2} \right)}} \right] - 2 \Lambda_I \left[ 1 - \frac{1 - \tanh{\left( \frac{b_I (R - R_I)}{2} \right)}}
            {1 + \tanh{\left( \frac{b_I R_I}{2} \right)}} \right],
        \end{equation}
    where $\Lambda$ is the current cosmological constant, while $\Lambda_I$ accounts for the effective cosmological constant in the
    early Universe. In the following it will not be considered the last part, the one that accounts for inflation, in (\ref{iii4}).
    Thus, the second model we take into consideration will be the one given by
        \begin{equation}\label{iii5}
            F(R) = R - 2 \Lambda \left[ 1 - \frac{1 - \tanh{\left( \frac{b_0 (R - R_0)}{2} \right)}}{1 + \tanh{\left(
            \frac{b_0 R_0}{2} \right)}} \right].
        \end{equation}

    As already shown in the first model, analytical solutions for Eq.(\ref{ii14}) cannot be found for (\ref{iii5}). The procedure followed in order to solve
    the problem is the same one used in the previous subsection; i.e., a set of parameters will be chosen for model (\ref{iii5}), then Eq.(\ref{ii14}) will be solved
    numerically, and, finally, I will check whether the results are in accordance with (\ref{iii6}) or not.

    For the model (\ref{iii5}), I set
        \begin{equation}\label{iii9}
            R_0 = 10^{-66} eV^2, \quad b = 1.16 R_0^{-1}, \quad \Lambda = 7.93 \tilde{m}^2.
        \end{equation}
    Following the same steps as in the previous subsection, it is found that the initial conditions are given by
        \begin{equation}\label{iii10}
            \left. y_H(z) \right|_{z_i} = \frac{\Lambda}{3 \tilde{m}^2} \left( 1 - \frac{1 - \tanh{\left( b
            \frac{3 \tilde{m}^2 (1 + z)^3 - R_0}{2} \right)}}{1 + \tanh{\left( \frac{b R_0}{2} \right)}} \right),$$
            $$\left. \frac{dy_H(z)}{dz} \right|_{z_i} = \frac{3 b \Lambda}{2} \frac{\left[ \cosh{\left( b
            \frac{3 \tilde{m}^2 (1 + z)^3 - R_0}{2} \right)} \right]^{-2}}{1 + \tanh{\left( \frac{b R_0}{2} \right)}} (1+z)^2.
        \end{equation}
    And, finally, for model (\ref{iii5}) with the set of parameters (\ref{iii9}) and initial conditions given by (\ref{iii10}), I set $z_i = 2.51$, obtaining
    $\omega_{DE}(0) = -0.969$ and $\Omega_{DE} = 0.735$, which are also in accordance with the
    observational data given by (\ref{iii6}).

    From now on, model (\ref{ii5}) with the set of parameters given by (\ref{iii9}) will be called {\bf model II}.

%%%%%%%%%%%%%%%%%%%%%%%%%%%
%%%  Sec. IV
%%%%%%%%%%%%%%%%%%%%%%%%%%%
\section{Growth of matter perturbations: Growth rate}

    In this section, the growth of matter density perturbations, $\delta = \frac{\delta \rho_\mathrm{m}}{\rho_\mathrm{m}}$, for model I and model II is studied.
    Since it is known that many different gravitational theories can mimic the $\Lambda$CDM unverse, which is commonly accepted as the Universe in which we live,
    the study of the growth history of these theories may be considered an essential tool to discriminate among them.

    Under the subhorizon approximation, the matter density perturbation $\delta = \frac{\delta \rho_\mathrm{m}}{\rho_\mathrm{m}}$ satisfies the following equation
    \cite{Tsujikawa:2007gd} (and references therein):
        \begin{equation}\label{iv1}
            \ddot \delta \, + \, 2 H \dot \delta \, - \, 4 \pi G_\mathrm{eff}(a,k) \rho_\mathrm{m} \delta = 0,
        \end{equation}
    with $k$ being the comoving wave number and $G_\mathrm{eff}(a,k)$ being the effective gravitational ``constant'' given by
        \begin{equation}\label{iv2}
            G_\mathrm{eff}(a,k) = \frac{G}{F'(R)} \left[ 1 + \frac{\left( k^2/a^2 \right) \left( F''(R)/F'(R) \right)}{1 + 3
            \left( k^2/a^2 \right) \left( F''(R)/F'(R) \right)} \right]\,.
    \end{equation}

        %%%%%%%%%%%%%%%%%%
        %%%%% Fig. 1 %%%%%
        %%%%%%%%%%%%%%%%%%
\begin{figure}[ht]
    \subfigure[\, $G_{eff}/G$ (model I)]{\includegraphics[width=0.47\textwidth]{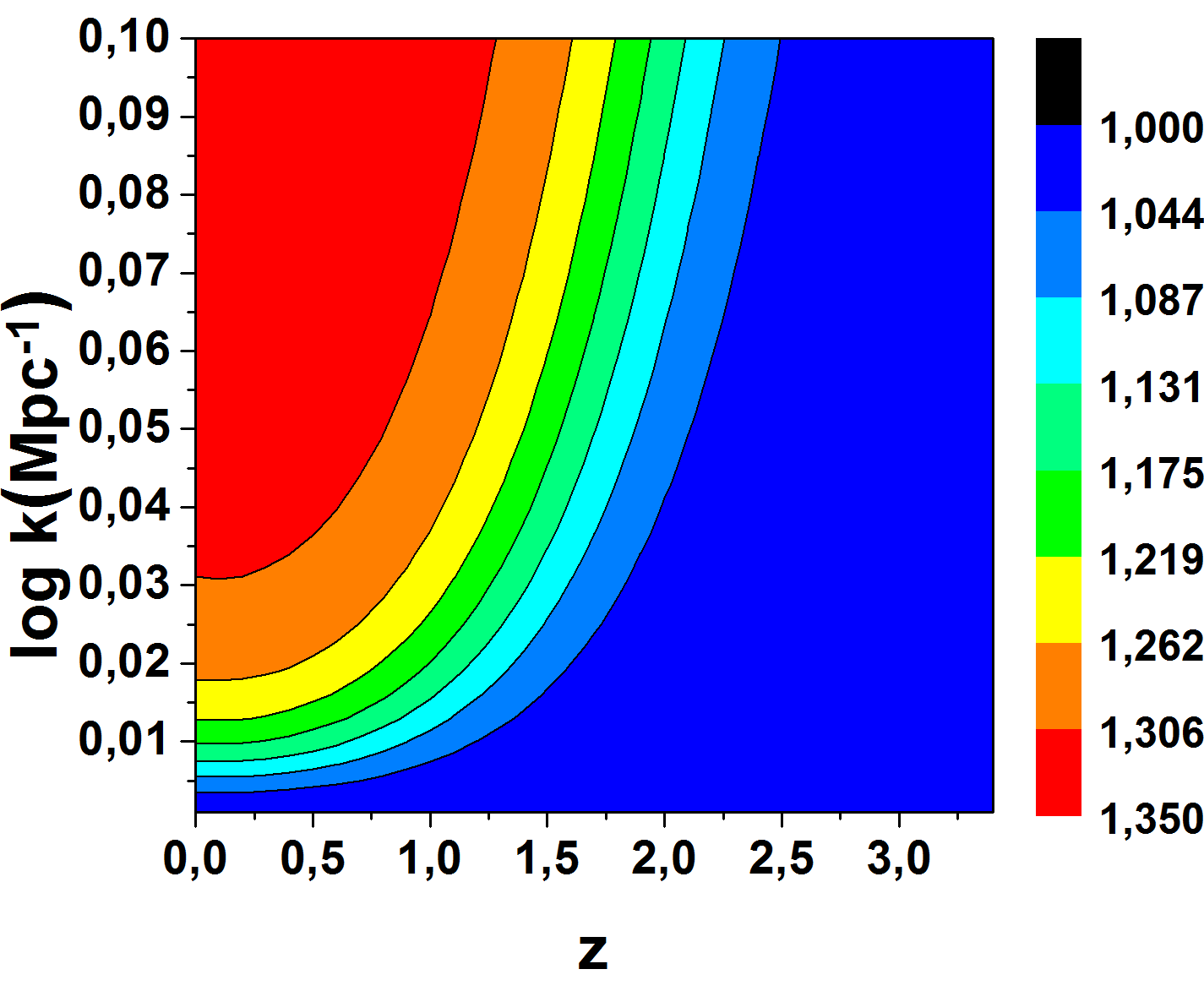}}
    \qquad
    \subfigure[\, $G_{eff}/G$ (model II)]{\includegraphics[width=0.47\textwidth]{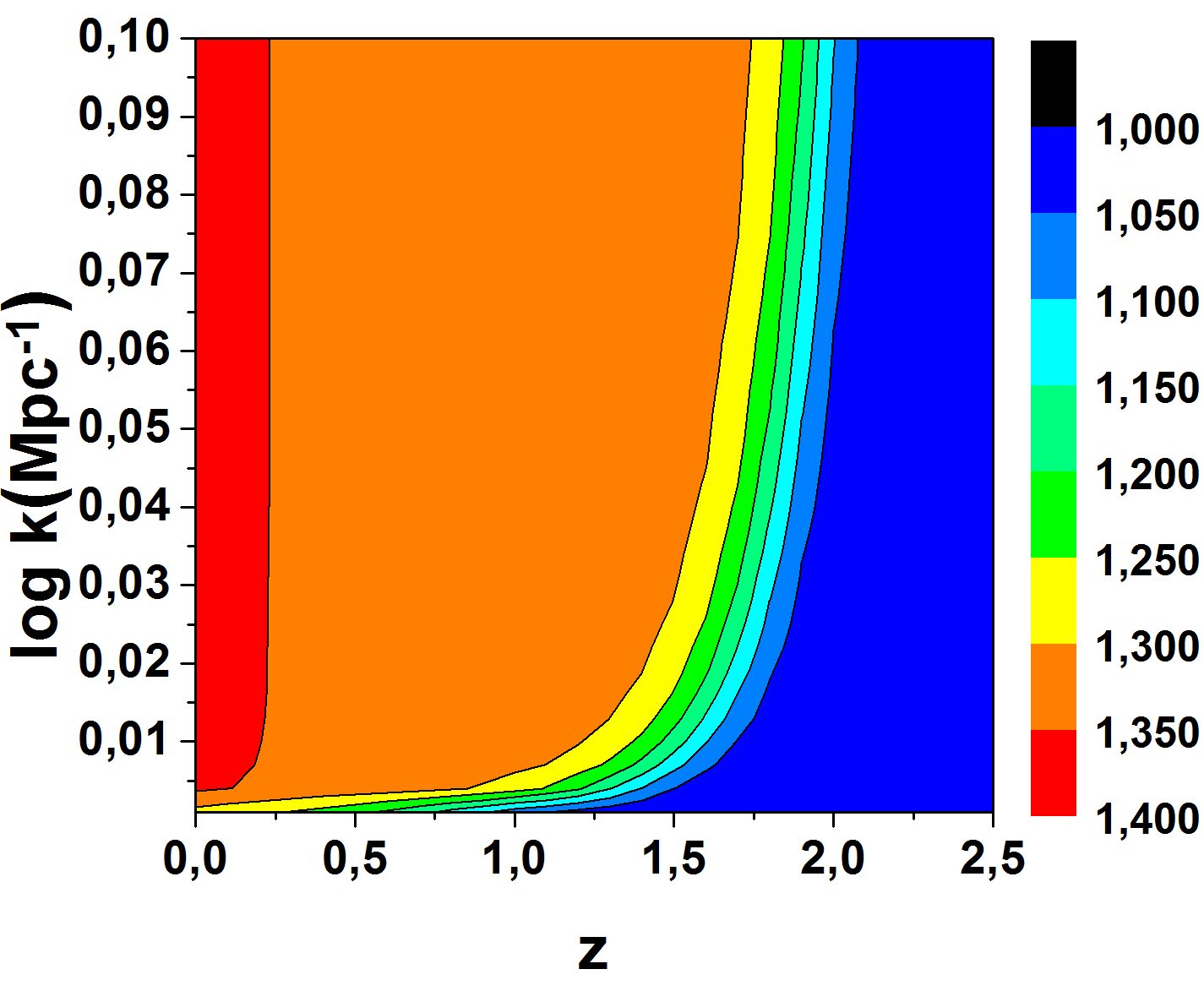}}
    \caption{Contour plot of the effective gravitational constant $G_\mathrm{eff}/G$ as a function of $z$ and $\log k (\mathrm{Mpc}^{-1})$ for model I (a)
    and model II (b).}
    \label{contour_eff_grav_const}
\end{figure}
%\begin{figure}[ht]
%    \subfigure[]{\includegraphics[width=0.47\textwidth]{MI_eff_grav_contour2}}
%    \qquad
%    \subfigure[]{\includegraphics[width=0.47\textwidth]{MI_eff_gravdef}}
%    \caption{(a) Contour plot of the effective gravitational constant $G_\mathrm{eff}/G$ as a function of redshift $z$ and the scale
%    dependence on the comoving wave number $k (\mathrm{Mpc}^{-1})$ for model I. The blue line, red and green lines indicate the sections depicted in
%    (b). The step for the contour lines is $0.015$. (b) Cosmological evolution of $G_\mathrm{eff}/G$ as a function of $z$ in model I for $k = 0.1
%    \mathrm{Mpc}^{-1}$ (blue line), $k = 0.01 \mathrm{Mpc}^{-1}$ (red) and $k = 0.001 \mathrm{Mpc}^{-1}$ (green line).}
%    \label{MI_eff_grav_const}
%\end{figure}
        %%%%%%%%%%%%%%%%%%
        %%%%%%%%%%%%%%%%%%
        %%%%% Fig. 2 %%%%%
        %%%%%%%%%%%%%%%%%%
\begin{figure}[!h]
    \subfigure[\, model I]{\includegraphics[width=0.45\textwidth]{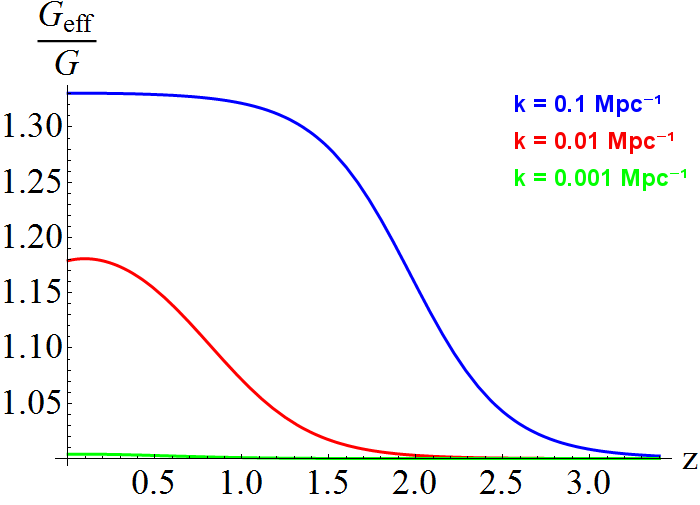}}
    \qquad
    \subfigure[\, model II]{\includegraphics[width=0.45\textwidth]{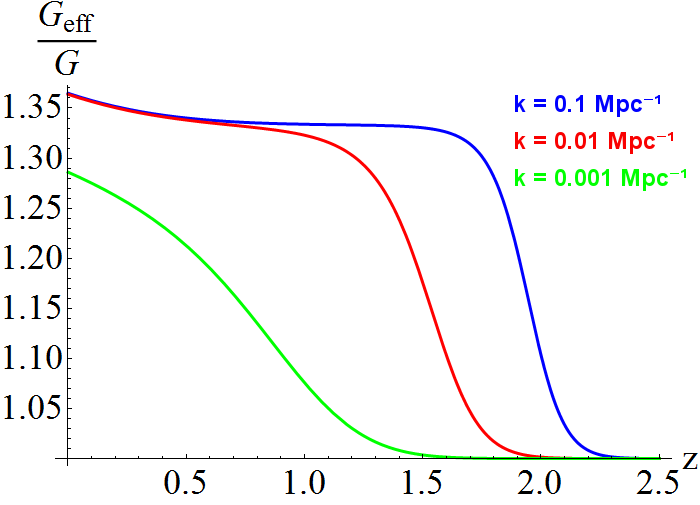}}
    \caption{Cosmological evolution of $G_\mathrm{eff}/G$ as a function of $z$ for $k = 0.1 \mathrm{Mpc}^{-1}$ (blue line), $k = 0.01 \mathrm{Mpc}^{-1}$ (red line) and
    $k = 0.001 \mathrm{Mpc}^{-1}$ (green line) for model I (a) and model II (b).}
    \label{eff_grav_const}
\end{figure}
%\begin{figure}[!h]
%    \subfigure[]{\includegraphics[width=0.45\textwidth]{MII_eff_grav_contour2}}
%    \qquad
%    \subfigure[]{\includegraphics[width=0.45\textwidth]{MII_eff_gravdef}}
%    \caption{(a) Contour plot of the effective gravitational constant $G_\mathrm{eff}/G$ as a function of redshift $z$ and the scale
%    dependence on the comoving wave number $k (\mathrm{Mpc}^{-1})$ for model II. The blue line, red and green lines indicate the sections depicted in (b).
%    The step for the contour lines is $0.017$. (b) Cosmological evolution of $G_\mathrm{eff}/G$ as a function of $z$ in model II for $k = 0.1 \mathrm{Mpc}^{-1}$
%    (blue line), $k = 0.01 \mathrm{Mpc}^{-1}$ (red) and $k = 0.001 \mathrm{Mpc}^{-1}$ (green line).}
%    \label{MII_eff_grav_const}
%\end{figure}
       %%%%%%%%%%%%%%%%%%
       %%%%%%%%%%%%%%%%%%
       %%%%% Fig. 3 %%%%%
       %%%%%%%%%%%%%%%%%%
\begin{figure}[!h]
    \subfigure[\, $f_\mathrm{g}$ (model I)]{\includegraphics[width=0.45\textwidth]{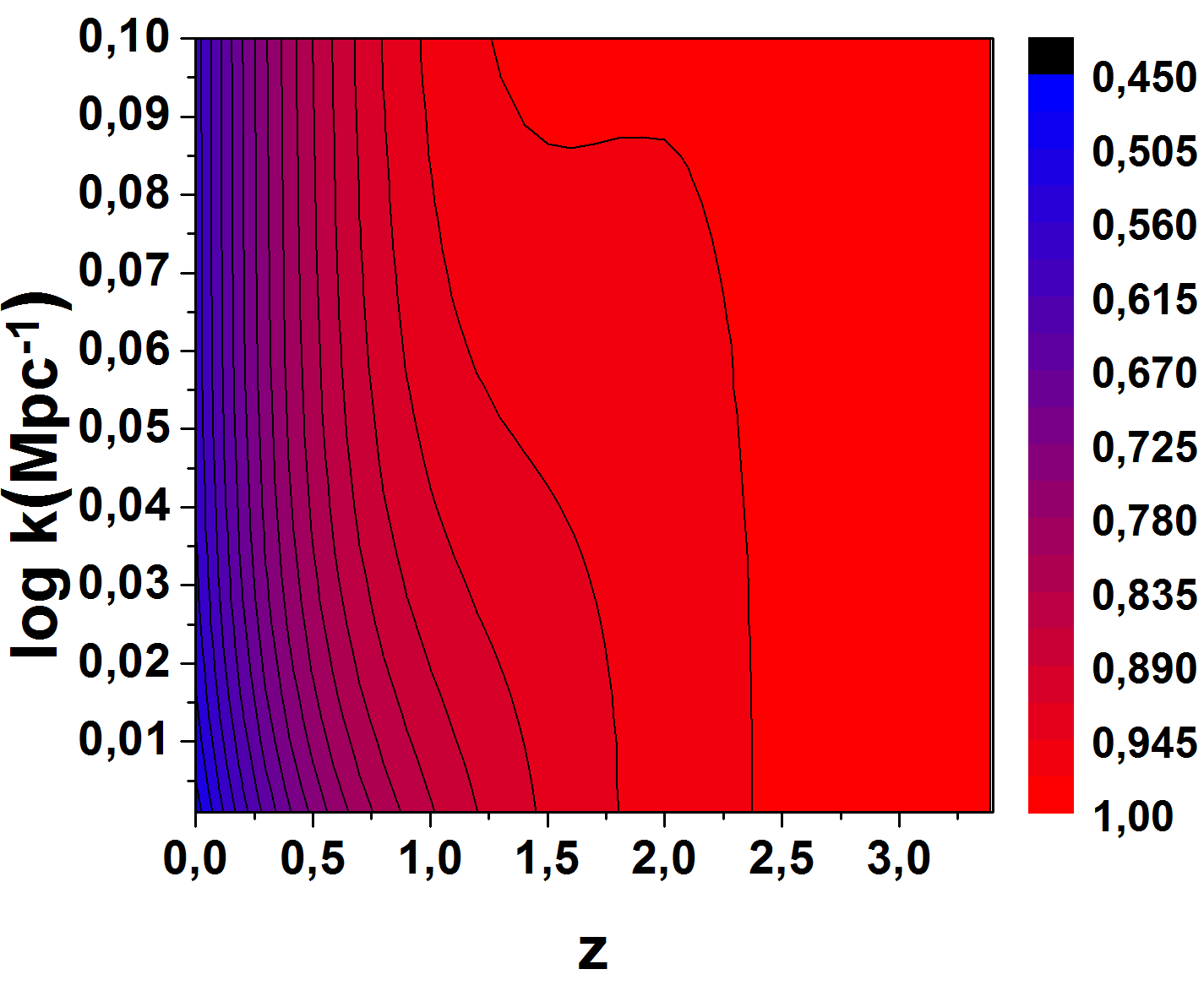}}
    \qquad
    \subfigure[\, $f_\mathrm{g}$ (model II)]{\includegraphics[width=0.45\textwidth]{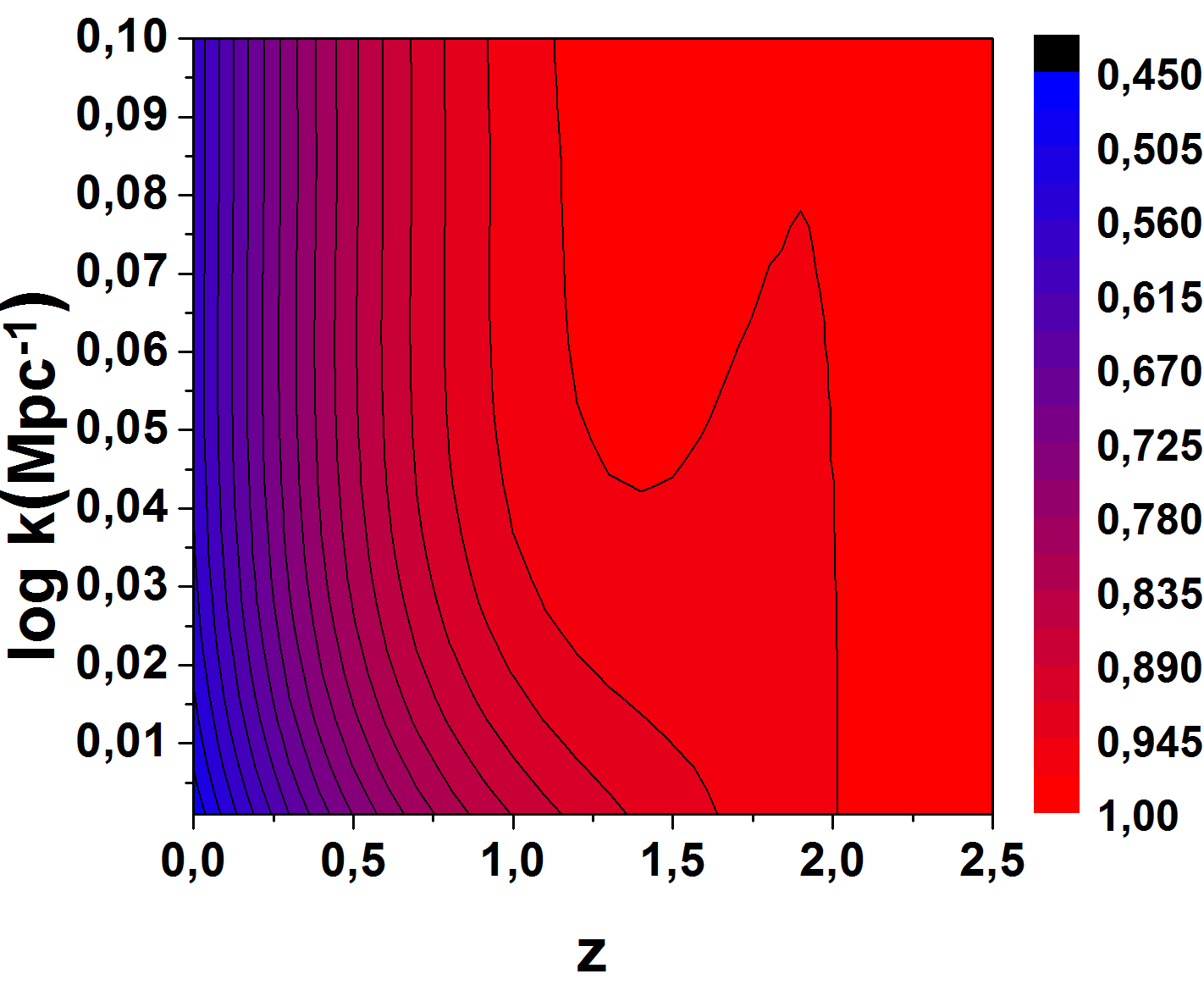}}
    \caption{Contour plot of the growth rate $f_\mathrm{g}$ as a function of $z$ and $\log k (\mathrm{Mpc}^{-1})$ for model I (a) and model II (b).}
    \label{contour_growth_rate}
\end{figure}
%\begin{figure}[!h]
%    \subfigure[]{\includegraphics[width=0.45\textwidth]{MI_growth_rate_contour2}}
%    \qquad
%    \subfigure[]{\includegraphics[width=0.45\textwidth]{MI_growth_ratedef}}
%    \caption{(a) Contour plot of the growth rate $f_\mathrm{g}$ as a function of the redshift $z$ and the scale dependence on the
%    comoving wave number $k (\mathrm{Mpc}^{-1})$ of the for model I. The blue line, red and green lines indicate the sections depicted in (b). The step for the
%    contour lines is $0.018$. (b) Cosmological evolution of the growth rate $f_\mathrm{g}$ as a function of $z$ in model I for $k = 0.1 \mathrm{Mpc}^{-1}$ (blue line),
%    $k = 0.01 \mathrm{Mpc}^{-1}$ (red) and $k = 0.001 \mathrm{Mpc}^{-1}$ (green line).}
%    \label{MI_growth_rate}
%\end{figure}
        %%%%%%%%%%%%%%%%%%
        %%%%%%%%%%%%%%%%%%
        %%%%% Fig. 4 %%%%%
        %%%%%%%%%%%%%%%%%%
\begin{figure}[!h]
    \subfigure[\, model I]{\includegraphics[width=0.45\textwidth]{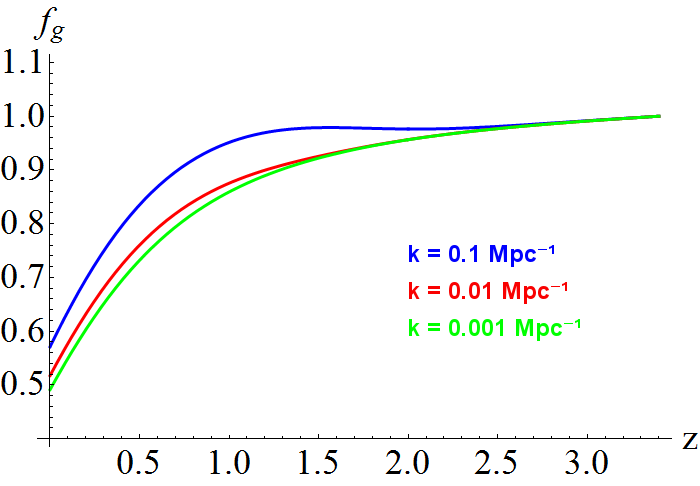}}
    \qquad
    \subfigure[\, model II]{\includegraphics[width=0.45\textwidth]{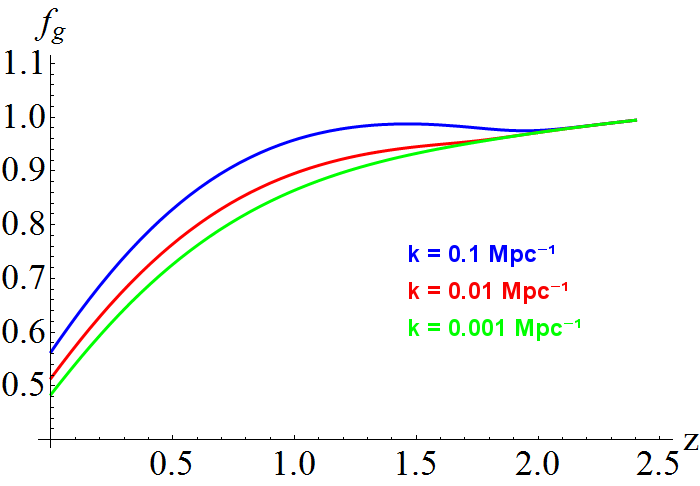}}
    \caption{Cosmological evolution of the growth rate $f_\mathrm{g}$ as a function of $z$ for $k = 0.1 \mathrm{Mpc}^{-1}$ (blue line),
    $k = 0.01 \mathrm{Mpc}^{-1}$ (red line) and $k = 0.001 \mathrm{Mpc}^{-1}$ (green line) for model I (a) and model II (b).}
\label{growth_rate}
\end{figure}
%\begin{figure}[!h]
%    \subfigure[]{\includegraphics[width=0.45\textwidth]{MII_growth_rate_contour2}}
%    \qquad
%    \subfigure[]{\includegraphics[width=0.45\textwidth]{MII_growth_ratedef}}
%    \caption{(a) Contour plot of the growth rate $f_\mathrm{g}$ as a function of the redshift $z$ and the scale dependence on the
%    comoving wave number $k (\mathrm{Mpc}^{-1})$ of the for model I. The blue line, red and green lines indicate the sections depicted in (b). The step for the
%    contour lines is $0.018$. (b) Cosmological evolution of the growth rate $f_\mathrm{g}$ as a function of $z$ in model II for $k = 0.1 \mathrm{Mpc}^{-1}$ (blue line),
%    $k = 0.01 \mathrm{Mpc}^{-1}$ (red) and $k = 0.001 \mathrm{Mpc}^{-1}$ (green line).}
%\label{MII_growth_rate}
%\end{figure}
        %%%%%%%%%%%%%%%%%%%

    In Figs.~\ref{contour_eff_grav_const} and \ref{eff_grav_const}, the cosmological evolution of the ratio $G_{eff}/G$ as a function of redshift $z$ and
    the comoving wave number $k$ for both model I and model II is depicted.

    The appearance of the comoving wave number $k$ in the expression of the effective gravitational constant $G_\mathrm{eff}$ has a huge importance due to the fact
    that now the evolution of the matter density perturbations also depends on $k$. This kind of dependence does not appear in the framework of general relativity.
    This fact can be easily checked by taking $F(R) = R$ in Eq.~(\ref{iv2}).

    In deriving Eq.~(\ref{iv1}), I assume the subhorizon approximation (see~\cite{EspositoFarese:2000ij}), for which the comoving wavelengths
    $\lambda \equiv a/k$ are considered to be much shorter than the Hubble radius $H^{-1}$. In terms of the comoving wave number, the subhorizon approximation can
    be written as follows:
        \begin{equation}\label{iv3}
            \frac{k^2}{a^2} \gg H^2\,.
        \end{equation}

    For model I, the subhorizon approximation states that $k \gg 0.000116 \mathrm{Mpc}^{-1}$. For model II, it states that $k \gg 0.000118 \mathrm{Mpc}^{-1}$. In order
    to fulfill (\ref{iv3}), the values considered for $k$ in this work will always satisfy $\log k \geq -3$, with $k$ written in $\mathrm{Mpc}^{-1}$, for both
    model I and model II. From now on, the expression $\log k(\mathrm{Mpc}^{-1})$ will be used to specify taking the logarithm of $k$, with $k$ written in
    $\mathrm{Mpc}^{-1}$. On the other hand, deviations from the linear regime have to be taken into account \cite{Cardone:2012xv} when $\log k(\mathrm{Mpc}^{-1})
    > -1$. Thus, the range of values considered for $k$ throughout this work for both model I and model II is $$-3 \leq \log k(\mathrm{Mpc}^{-1}) \leq -1 \, .$$

    Equation (\ref{iv1}) can be written in a different way by using the so-called growth rate $f_\mathrm{g}$ given by
    $f_\mathrm{g} \equiv d \ln{\delta}/d \ln{a}$. In terms of this growth rate, Eq.~(\ref{iv1}) reduces to
        \begin{equation}\label{iv4}
            \frac{df_\mathrm{g}(z)}{dz} \, + \, \left( \frac{1 + z}{H(z)} \frac{dH(z)}{dz} - 2 - f_\mathrm{g}(z) \right)
            \frac{f_\mathrm{g}(z)}{1 + z} + \frac{3}{2} \frac{\tilde{m}^2 (1 + z)^2}{H^2(z)} \frac{G_\mathrm{eff}(a(z),k)}{G} = 0\,.
        \end{equation}

    This equation can be solved numerically for model I and model II by imposing the condition that at high redshift the results for the $\Lambda$CDM universe are
    recovered. In Figs.~\ref{contour_growth_rate} and \ref{growth_rate}, the growth rate is shown as a function of the redshift $z$ and the comoving wave number
    $k$ for model I and model II.

    The next step should be to use the growth rate of these two models to compare them, but a new problem comes up when we face this task. Equation (\ref{iv4}) usually must
    be solved numerically because of its complexity. This means that, generally, we will not have an analytic expression for the growth rate to deal with. In order
    to compare and discriminate among different theories it would be helpful to have one or more parameters that characterize their growth history.

        %%%%%%%%%%%%%%%%%%
        %%%%% Fig. 5 %%%%%
        %%%%%%%%%%%%%%%%%%
%\begin{figure}
%  \centering
%    \includegraphics[width=0.7\textwidth]{constant_growth_index_vs_logk}
%  \caption{Constant growth index as a function of $\log k(Mpc^{-1})$ for model I (red) and for model II (blue line). The bars express the $68\%$ CL.}
%  \label{constant_growth_index_vs_logk}
%\end{figure}

\begin{figure}
    \subfigure[\ model I]{\includegraphics[width=0.49\textwidth]{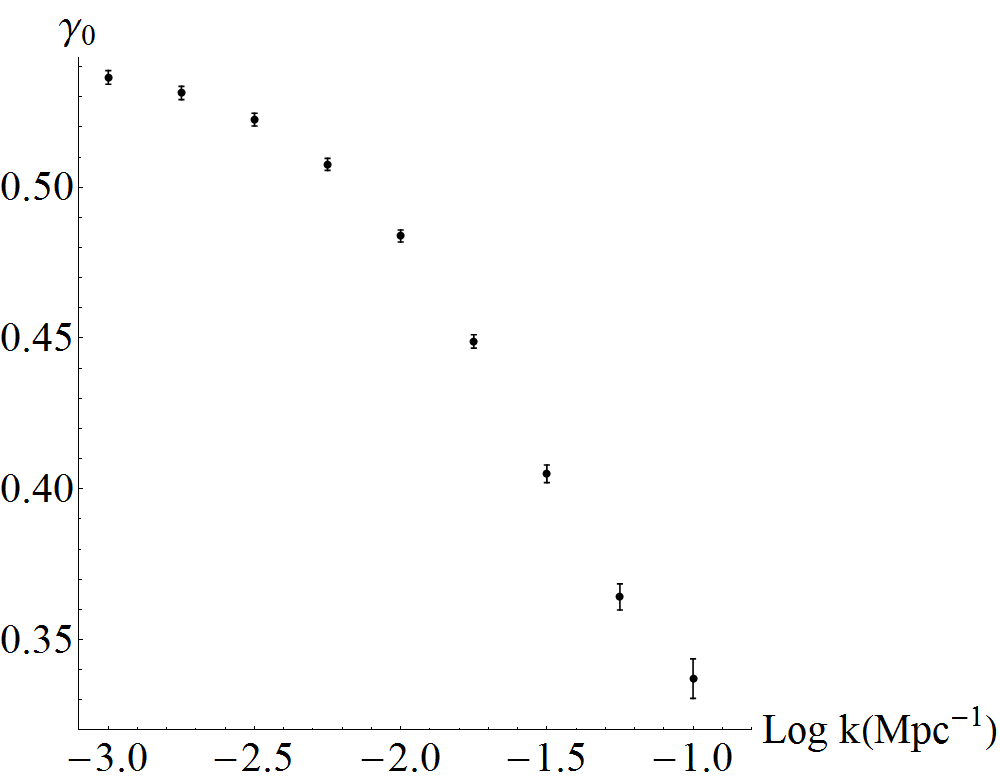}}
    \subfigure[\ model II]{\includegraphics[width=0.49\textwidth]{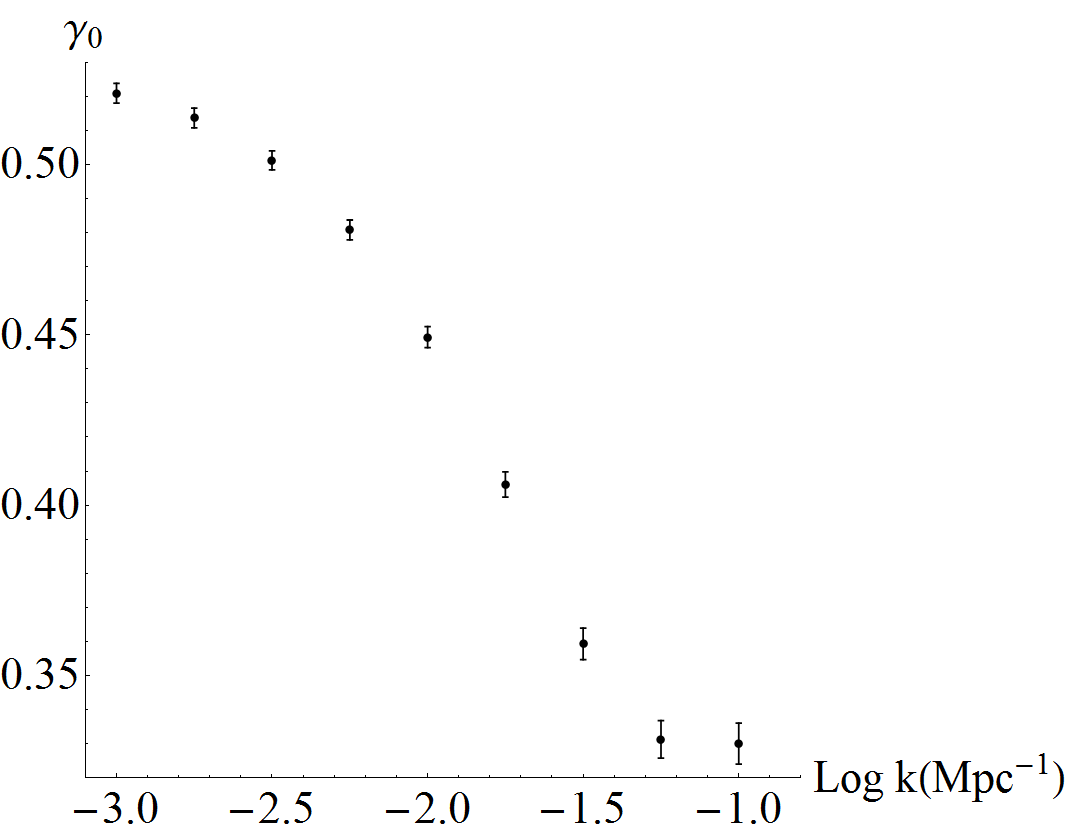}}
    \caption{Constant growth index as a function of $\log k(\mathrm{Mpc}^{-1})$ for model I (a) and for model II (b). The bars express the $68\%$ CL.}
    \label{constant_growth_index_vs_logk}
\end{figure}
        %%%%%%%%%%%%%%%%%%
        %%%%%%%%%%%%%%%%%%
        %%%%% Fig. 6 %%%%%
        %%%%%%%%%%%%%%%%%%
\begin{figure}[!h]
    \subfigure[\ model I]{\includegraphics[width=0.41\textwidth]{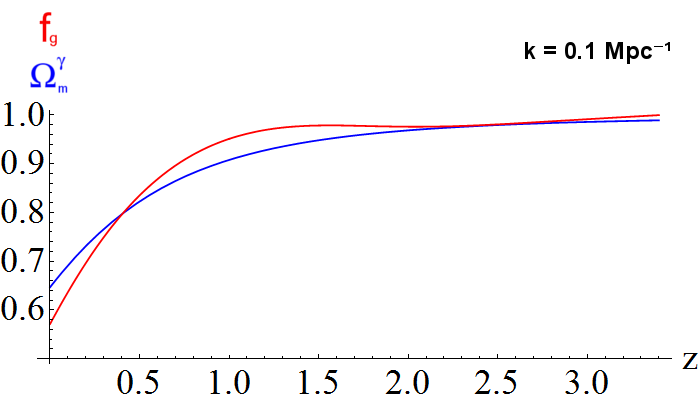}}
    \qquad
    \subfigure[\ model II]{\includegraphics[width=0.41\textwidth]{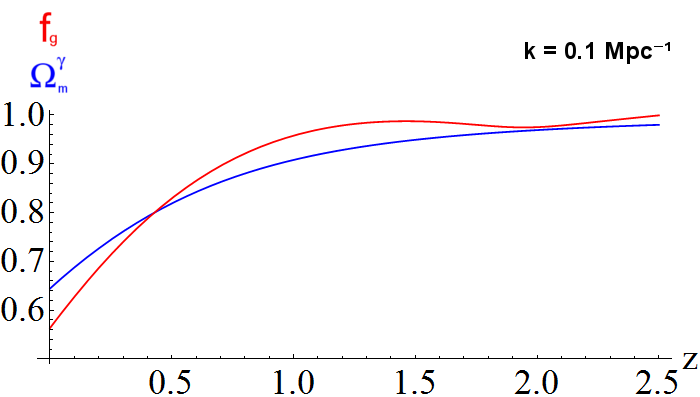}}
    \qquad
    \subfigure[\ model I]{\includegraphics[width=0.41\textwidth]{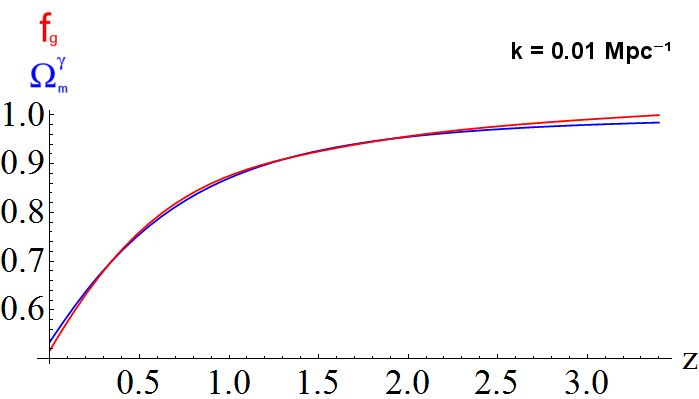}}
    \qquad
    \subfigure[\ model II]{\includegraphics[width=0.41\textwidth]{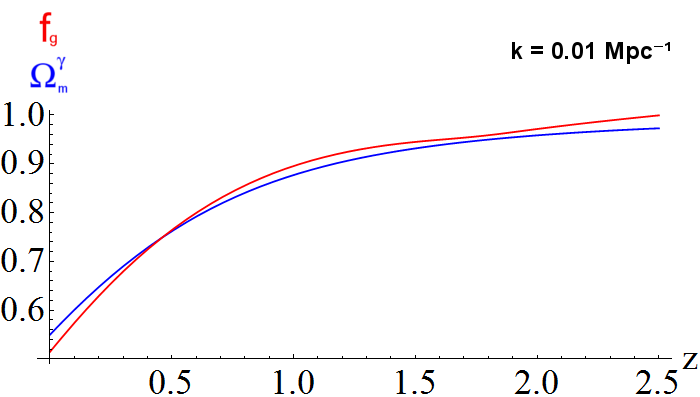}}
    \qquad
    \subfigure[\ model I]{\includegraphics[width=0.41\textwidth]{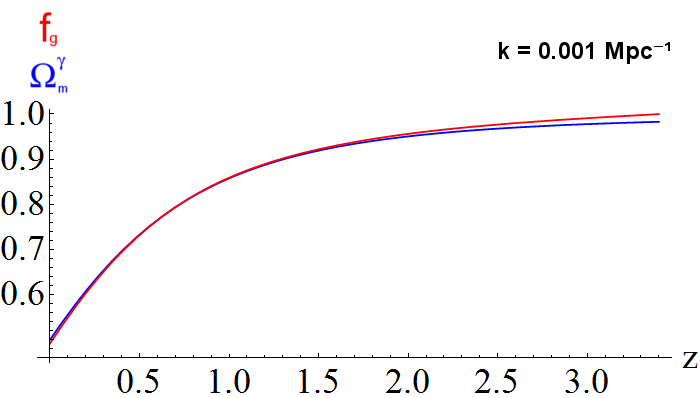}}
    \qquad
    \subfigure[\ model II]{\includegraphics[width=0.41\textwidth]{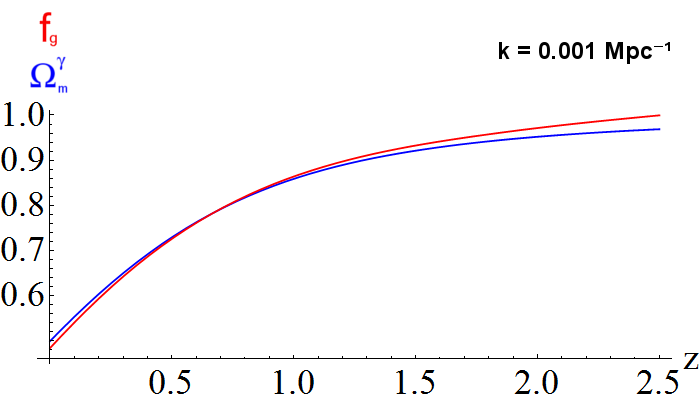}}
    \caption{Cosmological evolutions of the growth rate $f_\mathrm{g}$ (red line) and $\Omega_\mathrm{m}^\gamma$ (blue line) with $\gamma = \gamma_0$
    as functions of the redshift $z$ in model I for $k = 0.1 \mathrm{Mpc}^{-1}$ (a), $k = 0.01 \mathrm{Mpc}^{-1}$ (c) and $k = 0.001
    \mathrm{Mpc}^{-1}$ (e), and those in model II for $k = 0.1 \mathrm{Mpc}^{-1}$ (b), $k = 0.01 \mathrm{Mpc}^{-1}$ (d) and $k = 0.001
    \mathrm{Mpc}^{-1}$ (f).}
    \label{figure_constant_growth_index}
\end{figure}
        %%%%%%%%%%%%%%%%%%
        %%%%%%%%%%%%%%%%%%
        %%%%% Fig. 7 %%%%%
        %%%%%%%%%%%%%%%%%%
\begin{figure}[!h]
    \subfigure[\ model I]{\includegraphics[width=0.45\textwidth]{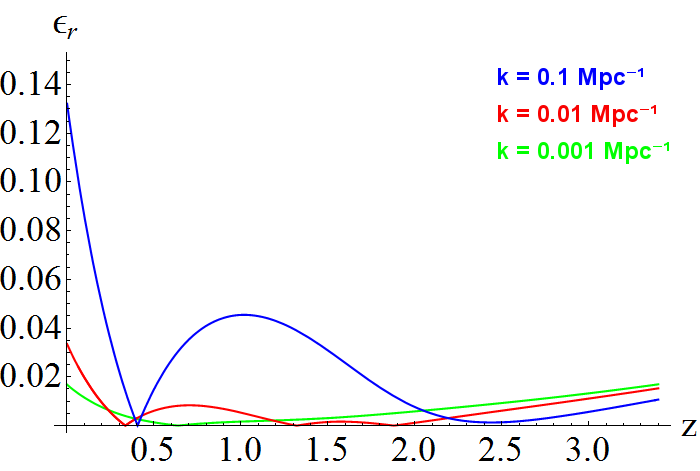}}
    \quad
    \subfigure[\ model II]{\includegraphics[width=0.45\textwidth]{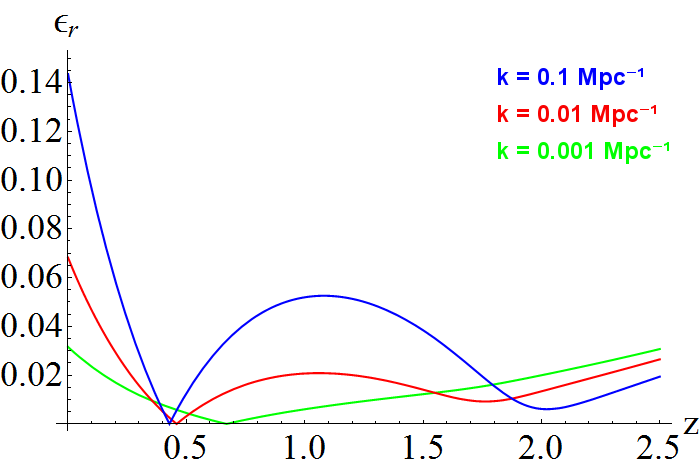}}
    \caption{Cosmological evolution of the relative difference $\epsilon_r = \frac{\left| f_\mathrm{g} - \Omega_\mathrm{m}^\gamma \right|}{f_\mathrm{g}}$ with
    $\gamma = \gamma_0$ for $k = 0.1 \mathrm{Mpc}^{-1}$ (blue line), $k = 0.01 \mathrm{Mpc}^{-1}$ (red line) and $k = 0.001 \mathrm{Mpc}^{-1}$ (green line) in
    model I (a) and model II (b).}
    \label{const_rel_dif}
\end{figure}
        %%%%%%%%%%%%%%%%%%%

%%%%%%%%%%%%%%%%%%%%%%%%%%%
%%%  Sec. V
%%%%%%%%%%%%%%%%%%%%%%%%%%%
\section{Characterizing the growth history: Growth index}

    In this section the concept of the growth index is developed. The growth index $\gamma$ appears as an important quantity in characterizing the growth of matter
    density perturbations.

    In order to compare the growth of matter density perturbations between different theories, the so-called growth index $\gamma$ appears. This index is given by
        \begin{equation}\label{iv5}
            f_\mathrm{g}(z) = \Omega_\mathrm{m}(z)^{\gamma(z)},
        \end{equation}
    where $\Omega_\mathrm{m}(z) = \frac{8 \pi G \rho_m}{3 H^2}$ is the matter density parameter. The growth index $\gamma$ cannot be directly observed, but it
    could have a huge importance in discriminating among different gravitational theories; it can be inferred from the observational data of both the growth
    factor $f_\mathrm{g}(z)$ and the matter density parameter $\Omega_\mathrm{m}(z)$ at the same redshift $z$.

    As it was done in \cite{Bamba:2012qi}, different parametrizations for the growth index $\gamma$ will be considered for both model I and model II. In a first
    stage, a constant $\gamma$ will be assumed \cite{Kaiser:1987qv, Hamilton:1997zq}; afterwards, a linear dependence \cite{Polarski:2007rr} given by
    $\gamma(z) = \gamma_0 + \gamma'_0 \cdot z$ will be considered; and, finally, an ansatz of the type $\gamma(z) = \gamma_0 + \gamma_1 \cdot z/(1 + z)$ will be
    suggested.

    In what follows, we will study these different parametrizations of the growth index for model I and model II.

    \subsection{$\gamma = \gamma_0$}

        We consider the ansatz for the growth index given by
            \begin{equation}
                \gamma = \gamma_0,
            \end{equation}
        where $\gamma_0$ is a constant.

        The results obtained by fitting Eq.~(\ref{iv5}) to the solution of Eq.~(\ref{iv4}) for different values of the comoving wave number $k$ for model I and
        model II are shown in Fig.~\ref{constant_growth_index_vs_logk}, where the points denote the median value while the bars express the $68\%$ confidence
        level (CL). It can be easily observed that both models exhibit a strong and quite similar dependence on $\log k(\mathrm{Mpc}^{-1})$. Moreover, $\gamma$ seems
        to be worse determined for model II.

        The cosmological evolutions of the growth rate $f_\mathrm{g}(z)$ and the expression $\Omega_\mathrm{m}(z)^{\gamma_0}$ as functions of the redshift $z$
        for several values of the comoving wave number $k$ for model I and model II are depicted in Fig.~\ref{figure_constant_growth_index}. From what
        is shown in this figure it is clear that the worst fit is given for the highest value of the comoving wave number $k$.

        In order to clarify the results obtained, the relative difference between $f_\mathrm{g}(z)$ and $\Omega_\mathrm{m}(z)^{\gamma_0}$ is defined as
            \begin{equation}\label{relative}
                \epsilon_r (z,k) = \frac{\left| f_\mathrm{g}(z,k) - \Omega_\mathrm{m}(z)^\gamma \right|}{f_\mathrm{g}(z,k)}.
            \end{equation}
        In Fig.~\ref{const_rel_dif}, the cosmological evolution of $\epsilon_r$ as a function of $z$ for the same values of $k$ in both models are
        shown. For model I the relative difference is less than $13\%$ for $\log k(\mathrm{Mpc}^{-1}) = - 1$, $3.5\%$ for $\log k(\mathrm{Mpc}^{-1}) = -
        2$ and $2\%$ for $\log k(\mathrm{Mpc}^{-1}) = - 3$; while for model II, the highest value of $\epsilon_r$ is $14\%$ for $\log k(\mathrm{Mpc}^{-1}) = - 1$,
        $7\%$ for $\log k(\mathrm{Mpc}^{-1}) = - 2$ and $3\%$ for $\log k(\mathrm{Mpc}^{-1}) = - 3$. Thus, two points can be made. First of all, the
        fits for model I are, generally, better than the ones for model II. Second of all, the fits are better for lower values of
        $\log k(\mathrm{Mpc}^{-1})$ for both models.

        %%%%%%%%%%%%%%%%%%
        %%%%% Fig. 8 %%%%%
        %%%%%%%%%%%%%%%%%%
%\begin{figure}[!h]
%    \subfigure[]{\includegraphics[width=0.45\textwidth]{linear_growth_index_vs_logk_0}}
%    \quad
%    \subfigure[]{\includegraphics[width=0.45\textwidth]{linear_growth_index_vs_logk_1}}
%    \caption{ Growth index fitting parameters in the case $\gamma = \gamma_0 + \gamma_1 z$ as a function of $\log k(Mpc^{-1})$ for model I [(a) and (c)] and Model
%    II [(b) and (d)]. Legend is the same as Fig.~\ref{constant_growth_index_vs_logk}.}
%\label{linear_growth_index_vs_logk}
%\end{figure}

\begin{figure}[!h]
    \subfigure[\ model I]{\includegraphics[width=0.49\textwidth]{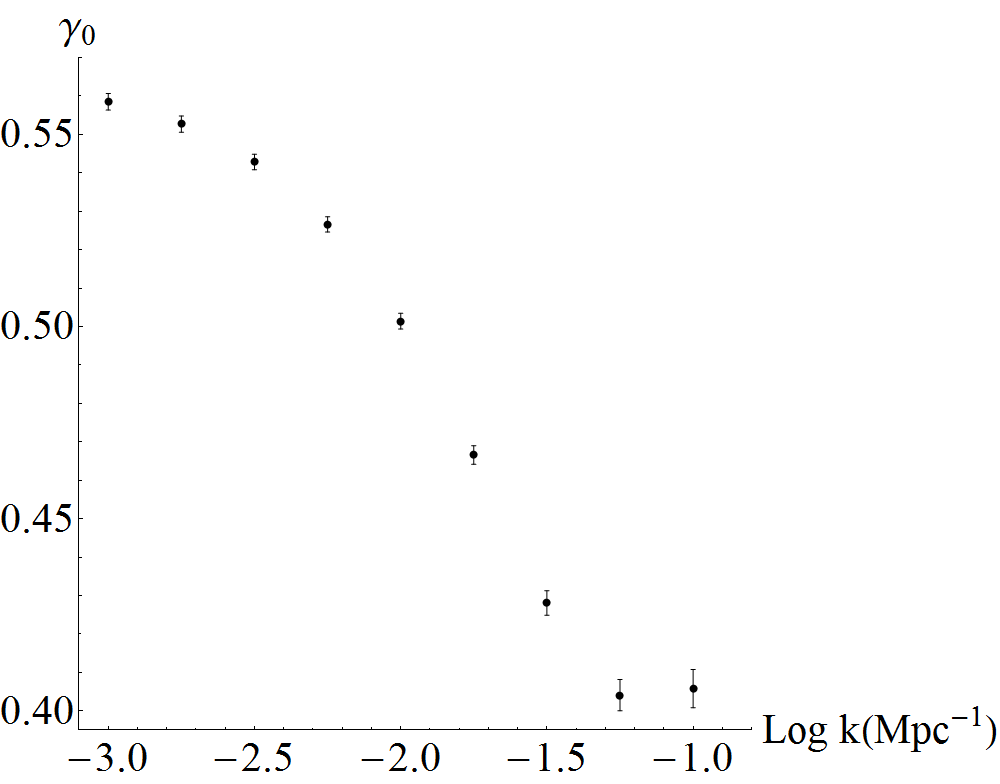}}
    \subfigure[\ model II]{\includegraphics[width=0.49\textwidth]{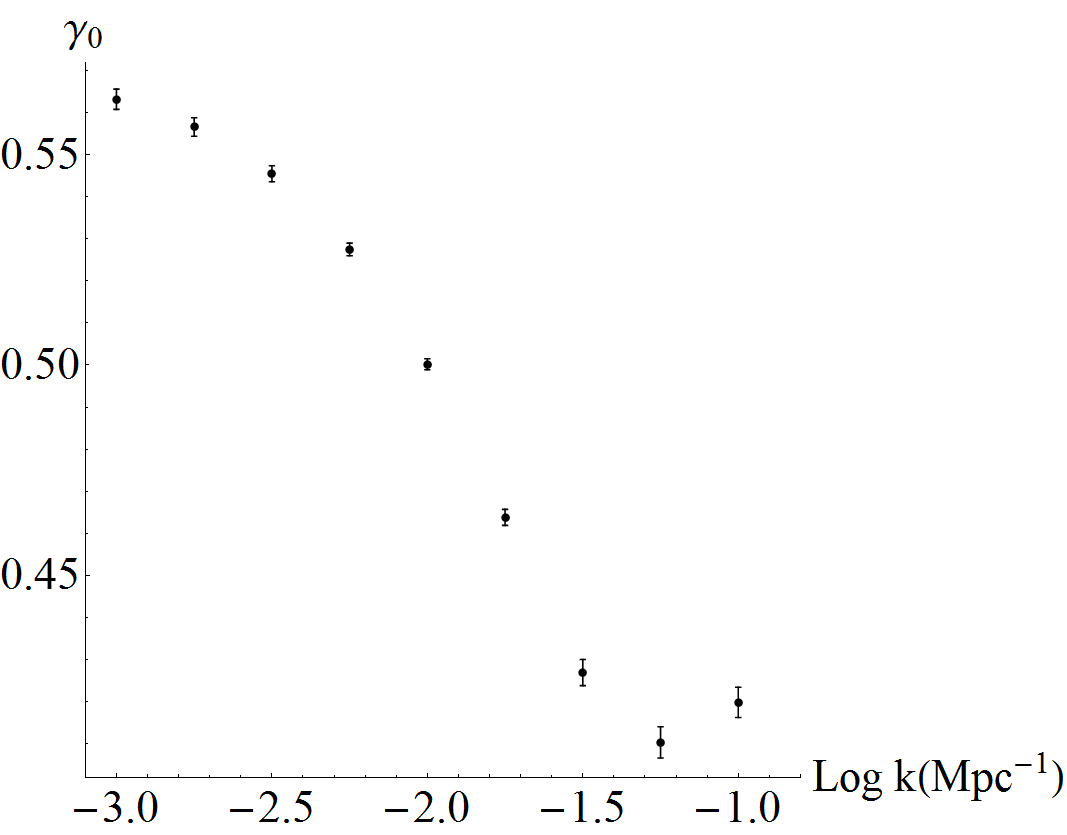}}
    \subfigure[\ model I]{\includegraphics[width=0.49\textwidth]{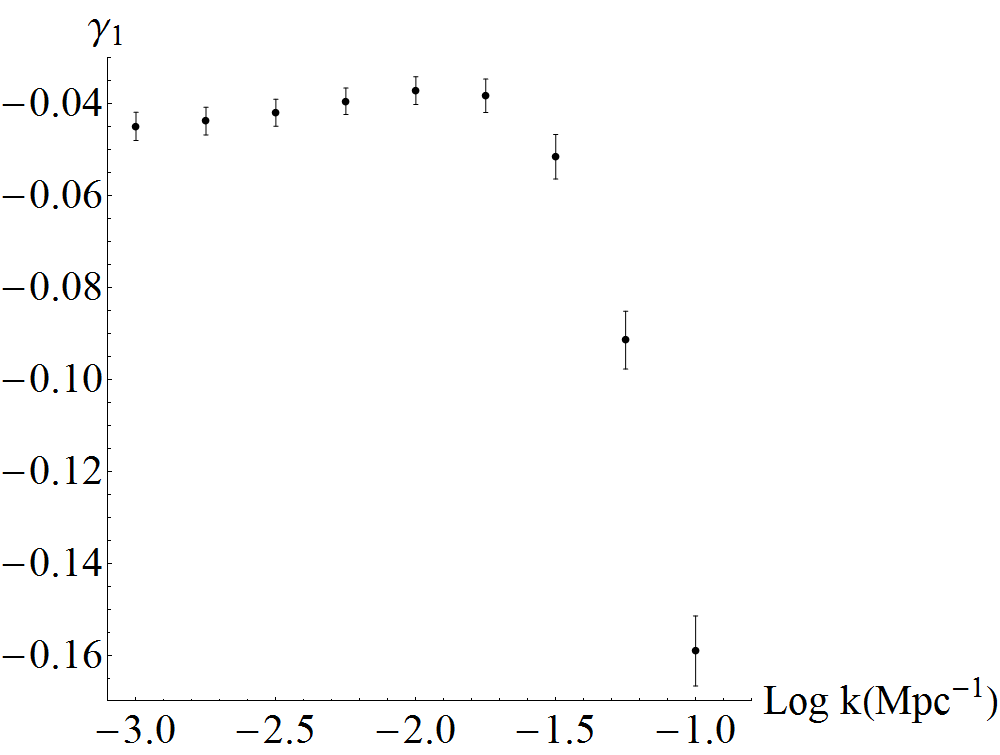}}
    \subfigure[\ model II]{\includegraphics[width=0.49\textwidth]{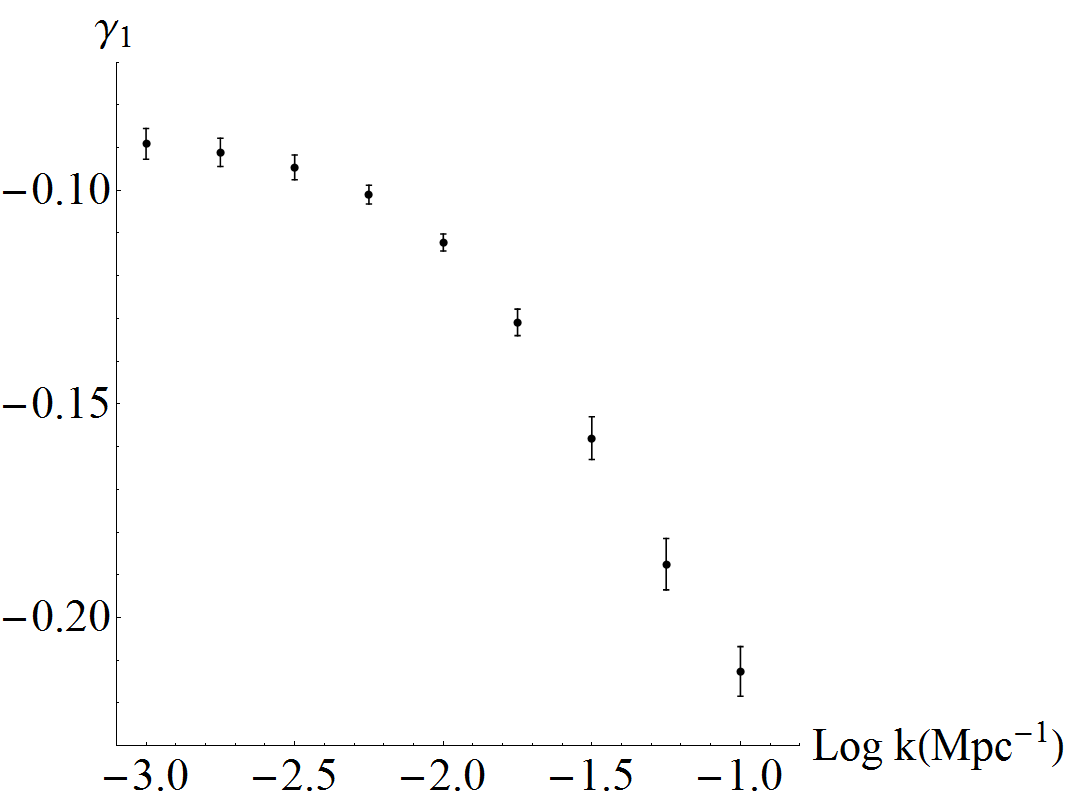}}
    \caption{ Growth index fitting parameters in the case $\gamma = \gamma_0 + \gamma_1 \cdot z$ as a function of $\log k(\mathrm{Mpc}^{-1})$ for model I [(a) and (c)]
    and model II [(b) and (d)]. The legend is the same as Fig.~\ref{constant_growth_index_vs_logk}.}
\label{linear_growth_index_vs_logk}
\end{figure}
        %%%%%%%%%%%%%%%%%%

    \subsection{$\gamma = \gamma_0 + \gamma_1 \cdot z$}

        In this subsection, the case of a growth index given by
            \begin{equation}
                \gamma = \gamma_0 + \gamma_1 \cdot z,
            \end{equation}
        where $\gamma_0$ and $\gamma_1$ are constants, will be studied following the same steps taken in the case of a constant growth index. The results obtained
        with this ansatz should improve those for a constant growth index.

        %%%%%%%%%%%%%%%%%%
        %%%%% Fig. 9 %%%%%
        %%%%%%%%%%%%%%%%%%
\begin{figure}[!h]
    \subfigure[\ model I]{\includegraphics[width=0.41\textwidth]{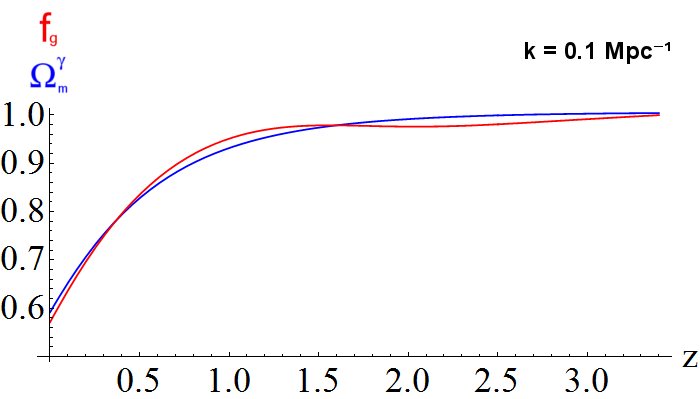}}
    \qquad
    \subfigure[\ model II]{\includegraphics[width=0.41\textwidth]{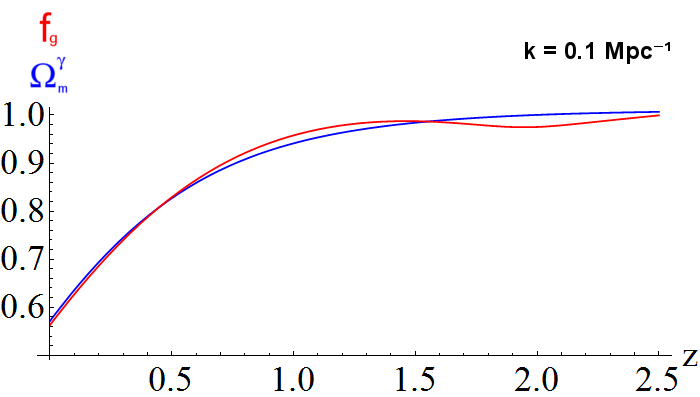}}
    \qquad
    \subfigure[\ model I]{\includegraphics[width=0.41\textwidth]{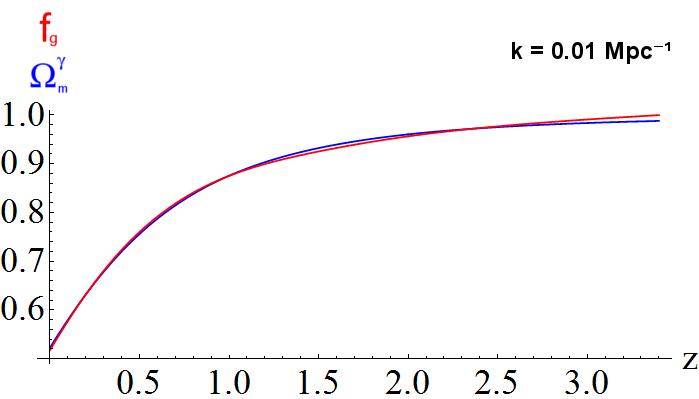}}
    \qquad
    \subfigure[\ model II]{\includegraphics[width=0.41\textwidth]{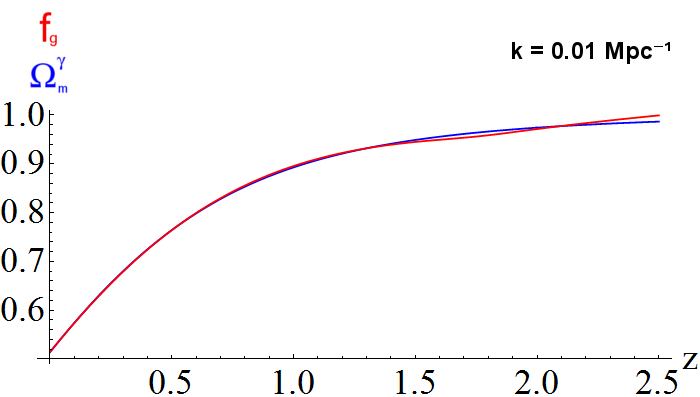}}
    \qquad
    \subfigure[\ model I]{\includegraphics[width=0.41\textwidth]{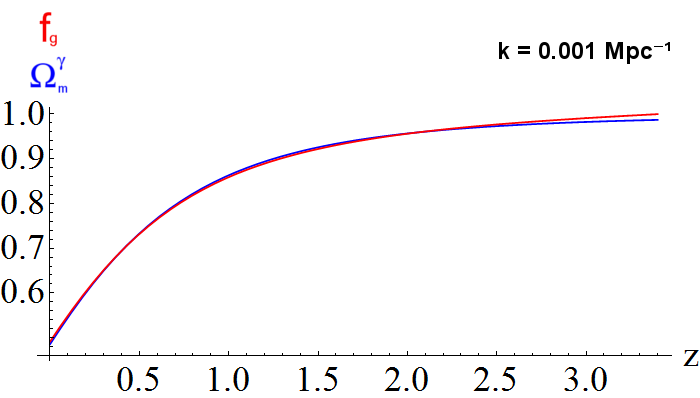}}
    \qquad
    \subfigure[\ model II]{\includegraphics[width=0.41\textwidth]{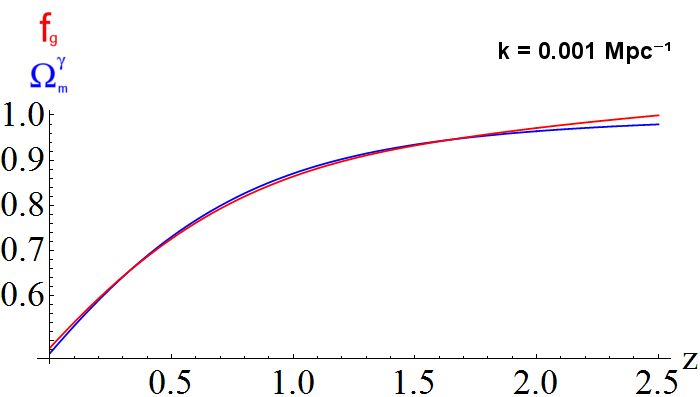}}
    \caption{Cosmological evolutions of the growth rate $f_\mathrm{g}$ (red line) and $\Omega_\mathrm{m}^\gamma$ (blue line) with $\gamma = \gamma_0
    + \gamma_1 \cdot z$ as functions of the redshift $z$ in model I for $k = 0.1 \mathrm{Mpc}^{-1}$ (a), $k = 0.01 \mathrm{Mpc}^{-1}$ (c) and $k
    = 0.001 \mathrm{Mpc}^{-1}$ (e), and those in model II for $k = 0.1 \mathrm{Mpc}^{-1}$ (b), $k = 0.01 \mathrm{Mpc}^{-1}$ (d) and $k =
    0.001 \mathrm{Mpc}^{-1}$ (f).}
    \label{figure_linear_growth_index}
\end{figure}
        %%%%%%%%%%%%%%%%%%%
        %%%%%%%%%%%%%%%%%%%
        %%%%% Fig. 10 %%%%%
        %%%%%%%%%%%%%%%%%%%
\begin{figure}[!h]
    \subfigure[\ model I]{\includegraphics[width=0.45\textwidth]{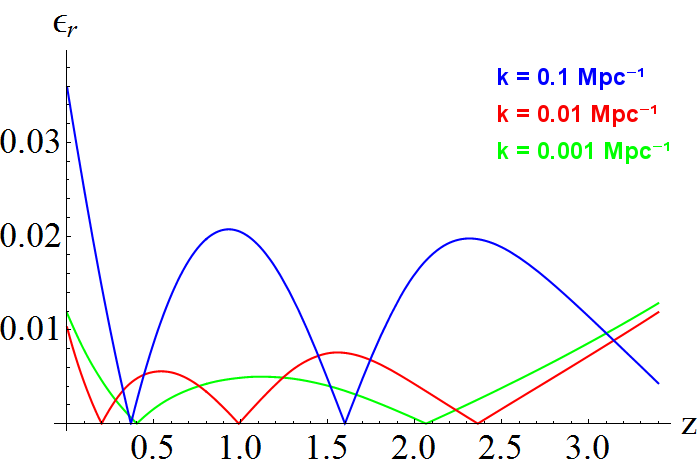}}
    \quad
    \subfigure[\ model II]{\includegraphics[width=0.45\textwidth]{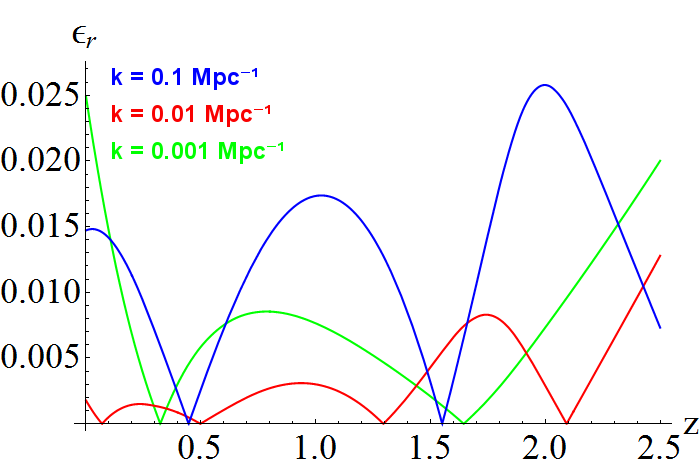}}
    \caption{Cosmological evolution of the relative difference $\epsilon_r = \frac{\left| f_\mathrm{g} - \Omega_\mathrm{m}^\gamma \right|}{f_\mathrm{g}}$
    with $\gamma = \gamma_0 + \gamma_1 \cdot z$ for $k = 0.1 \mathrm{Mpc}^{-1}$ (blue line), $k = 0.01 \mathrm{Mpc}^{-1}$ (red line) and $k = 0.001 \mathrm{Mpc}^{-1}$
    (green line) in model I (a) and model II (b).}
    \label{lin_rel_dif}
\end{figure}
        %%%%%%%%%%%%%%%%%%%

        In Fig.~\ref{linear_growth_index_vs_logk}, the parameters $\gamma_0$ and $\gamma_1$ for several values of $\log k(\mathrm{Mpc}^{-1})$ in both
        models are shown. For model I, $\gamma_0$ exhibits a clear dependence on $\log k(\mathrm{Mpc}^{-1})$, while $\gamma_1$ is almost constant for
        $-3 \leq \log k(\mathrm{Mpc}^{-1}) \leq -1.75$. For model II, the dependence on $\log k(\mathrm{Mpc}^{-1})$ is strong for both $\gamma_0$ and $\gamma_1$
        throughout the range of values considered for $\log k(\mathrm{Mpc}^{-1})$. It may also be noted that the parameter $\gamma_1$ gives the main difference
        between model I and model II.

        In Fig.~\ref{figure_linear_growth_index}, the cosmological evolutions of the growth rate $f_\mathrm{g}(z)$ and $\Omega_\mathrm{m}(z)^{\gamma(z)}$ as
        functions of the redshift $z$ together for model I and model II are depicted. Compared with the fits for a constant growth index, it can be easily
        noticed that the linear ansatz improves the results obtained, especially in the case of $\log k(\mathrm{Mpc}^{-1})=-1$.

        The cosmological evolution of the relative difference $\epsilon_r$ as a function of $z$ for several values of $k$ in model I and model II is shown in
        Fig.~\ref{lin_rel_dif}. For model I the relative difference is less than $3.5\%$ for $\log k(\mathrm{Mpc}^{-1}) = - 1$ and $1.5\%$ for
        $\log k(\mathrm{Mpc}^{-1}) \leq - 2$; while for model II, the highest value of $\epsilon_r$ is $2.5\%$ for $\log k(\mathrm{Mpc}^{-1}) = - 1$,
        $1.5\%$ for $\log k(\mathrm{Mpc}^{-1}) = - 2$ and $2.5\%$ for $\log k(\mathrm{Mpc}^{-1}) = - 3$. It might be accurate to note that these results improve
        those obtained for the case given by $\gamma = \gamma_0$, particulary those for the case $\log k(\mathrm{Mpc}^{-1}) = - 1$.

        %%%%%%%%%%%%%%%%%%%
        %%%%% Fig. 11 %%%%%
        %%%%%%%%%%%%%%%%%%%
\begin{figure}[!h]
    \subfigure[\ model I]{\includegraphics[width=0.49\textwidth]{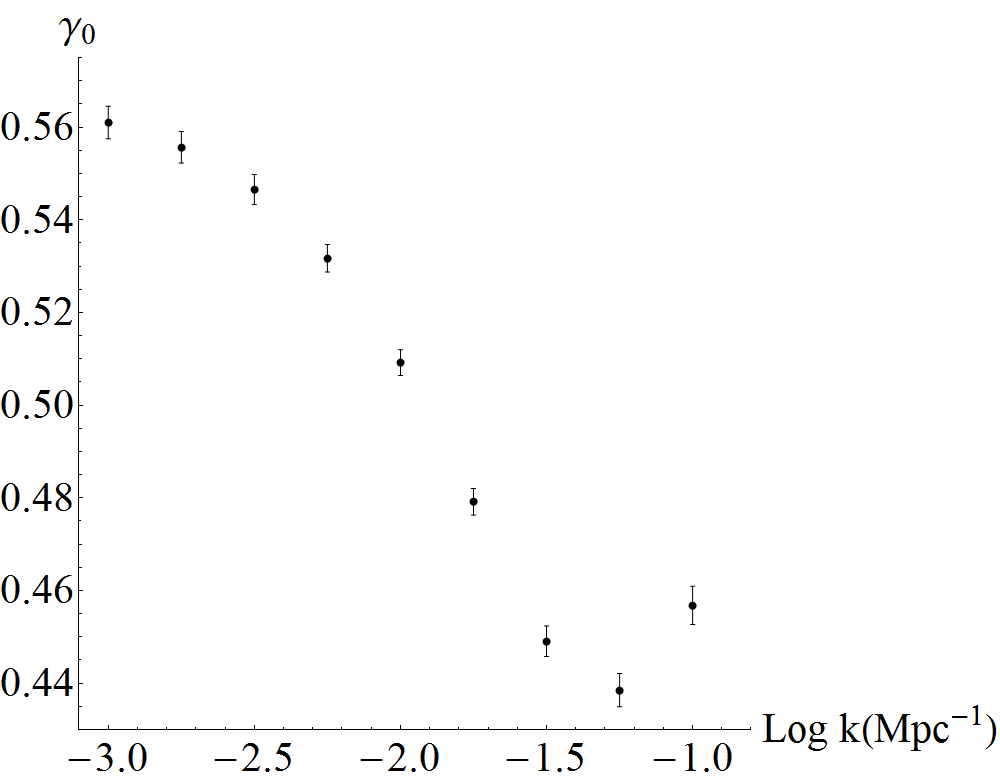}}
    \subfigure[\ model II]{\includegraphics[width=0.49\textwidth]{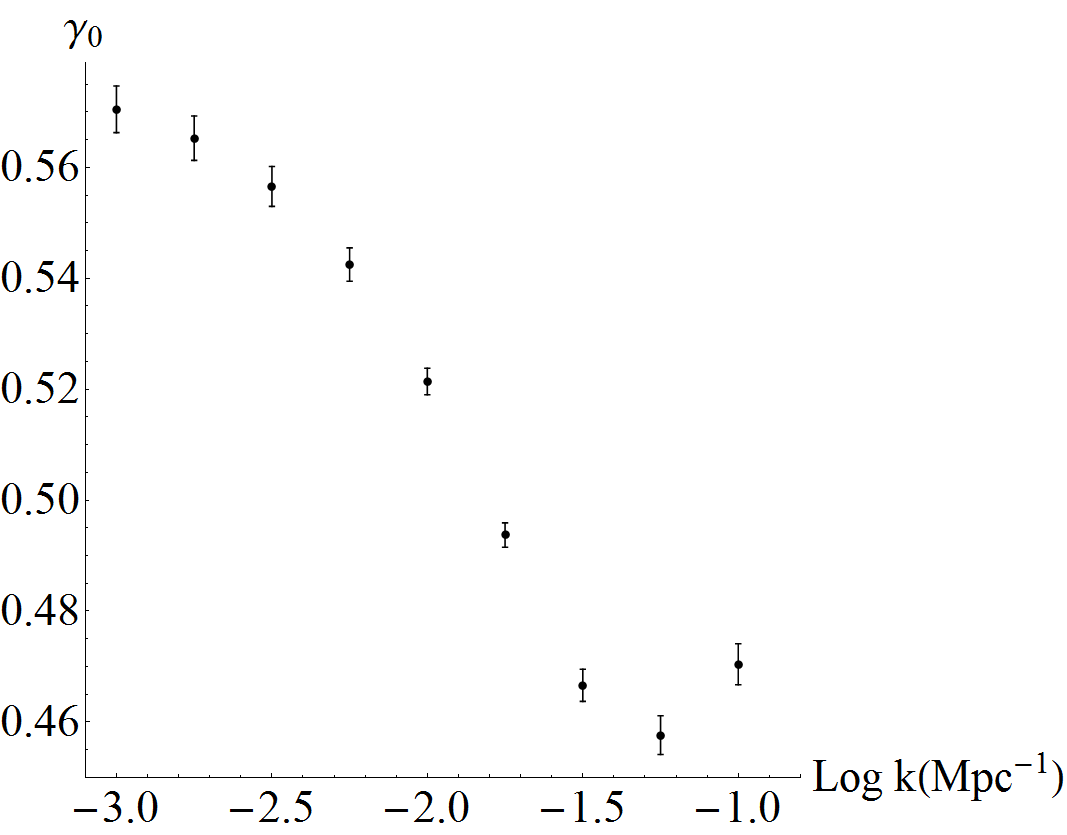}}
    \subfigure[\ model I]{\includegraphics[width=0.49\textwidth]{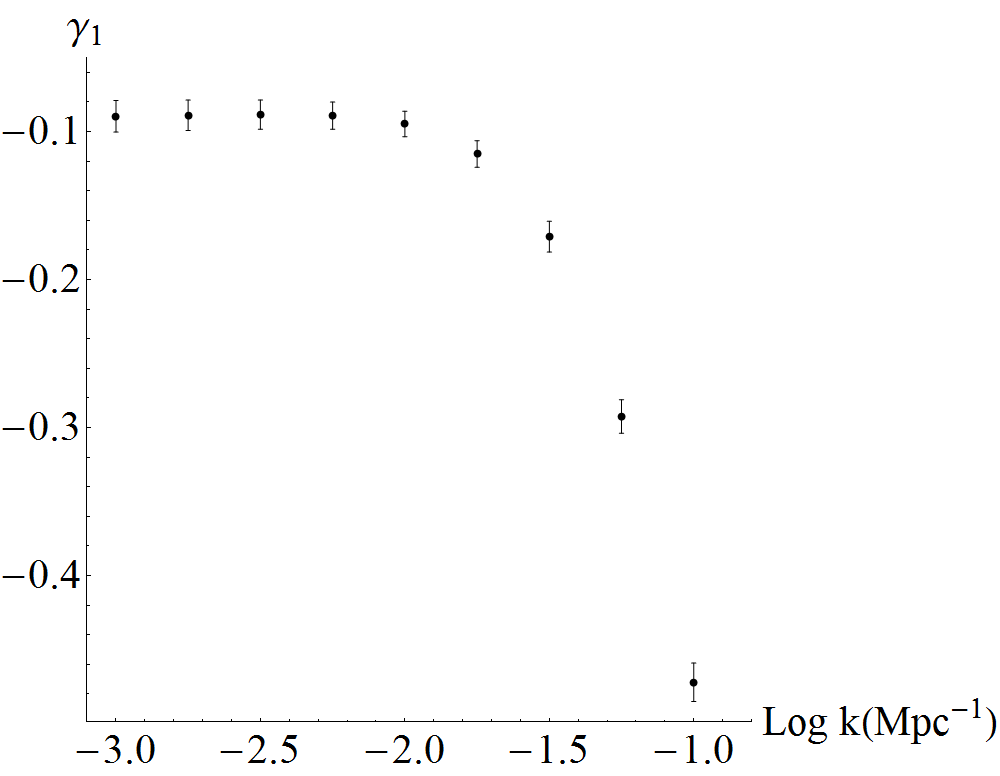}}
    \subfigure[\ model II]{\includegraphics[width=0.49\textwidth]{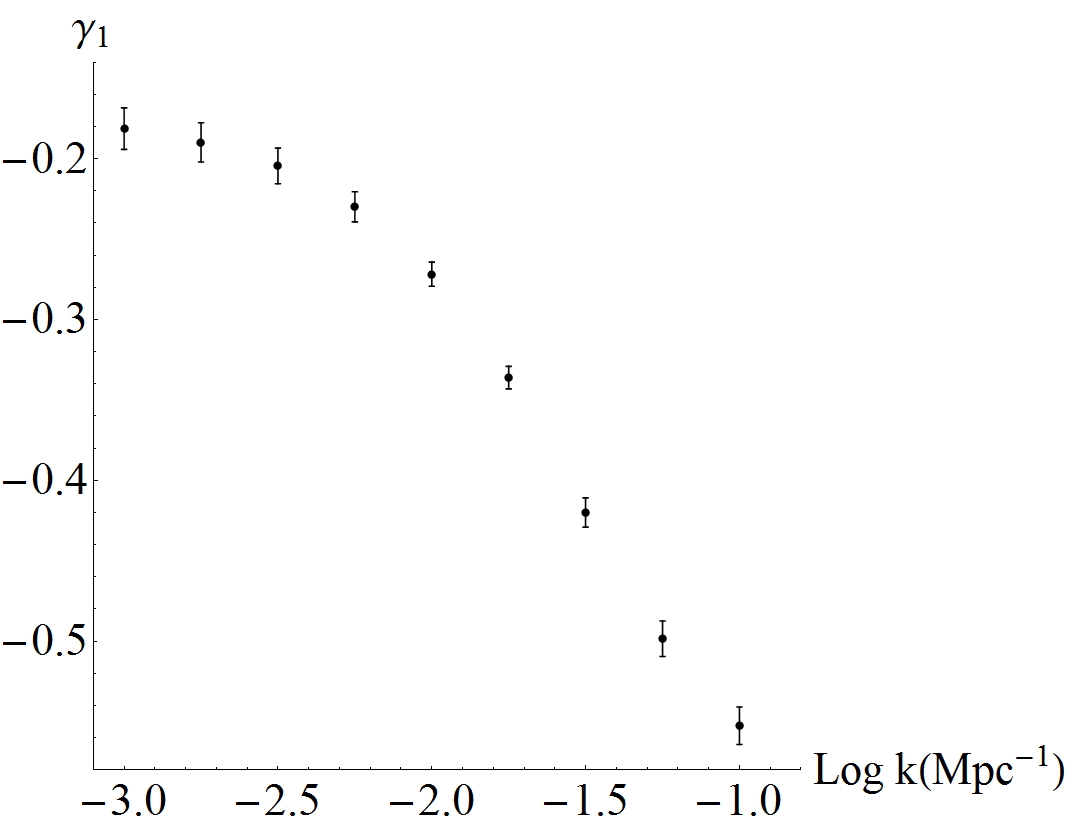}}
    \caption{Growth index fitting parameters in the case $\gamma = \gamma_0 + \gamma_1 \cdot \frac{z}{1 + z}$ as a function of $\log k(\mathrm{Mpc}^{-1})$ for
    model I [(a) and (c)] and model II [(b) and (d)]. The legend is the same as Fig.~\ref{constant_growth_index_vs_logk}. }
    \label{rational_growth_index_vs_logk}
\end{figure}
        %%%%%%%%%%%%%%%%%%%

    \subsection{$\gamma = \gamma_0 + \gamma_1 \cdot \frac{z}{1 + z}$}

        Finally, we assume the following ansatz for the growth index:
            \begin{equation}
                \gamma = \gamma_0 + \gamma_1 \cdot \frac{z}{1 + z},
            \end{equation}
        where $\gamma_0$ and $\gamma_1$ are constants.

        The parameters $\gamma_0$ and $\gamma_1$ for several values of $\log k$ are shown in Fig.~\ref{rational_growth_index_vs_logk} for both models. For model I,
        as it happened in the linear case, $\gamma_0$ exhibits a strong dependence on $\log k(\mathrm{Mpc}^{-1})$, while $\gamma_1$ is almost constant
        for $-3 \leq \log k(\mathrm{Mpc}^{-1}) \leq -2$. In the case of model II, the dependence on $\log k(\mathrm{Mpc}^{-1})$ is clear for both
        parameters $\gamma_0$ and $\gamma_1$.

        %%%%%%%%%%%%%%%%%%%
        %%%%% Fig. 12 %%%%%
        %%%%%%%%%%%%%%%%%%%
\begin{figure}[!h]
    \subfigure[\ model I]{\includegraphics[width=0.41\textwidth]{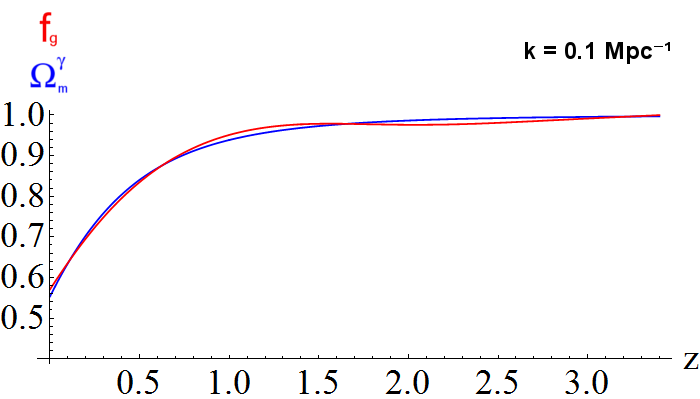}}
    \qquad
    \subfigure[\ model II]{\includegraphics[width=0.41\textwidth]{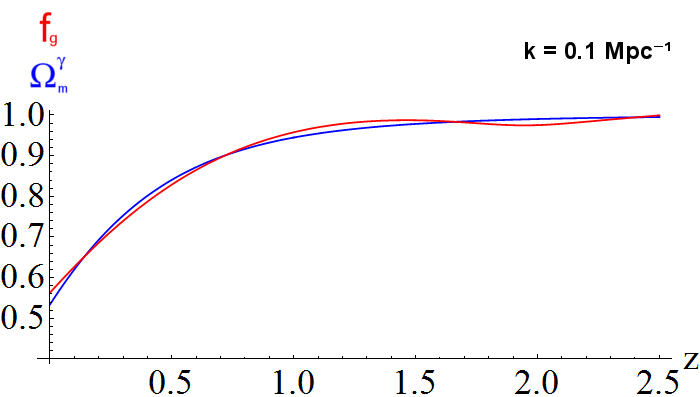}}
    \qquad
    \subfigure[\ model I]{\includegraphics[width=0.41\textwidth]{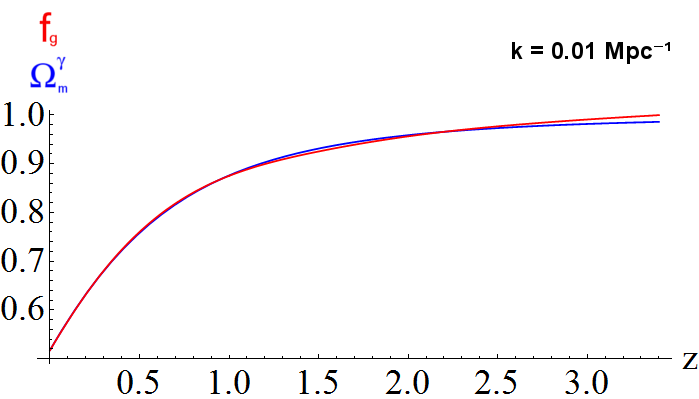}}
    \qquad
    \subfigure[\ model II]{\includegraphics[width=0.41\textwidth]{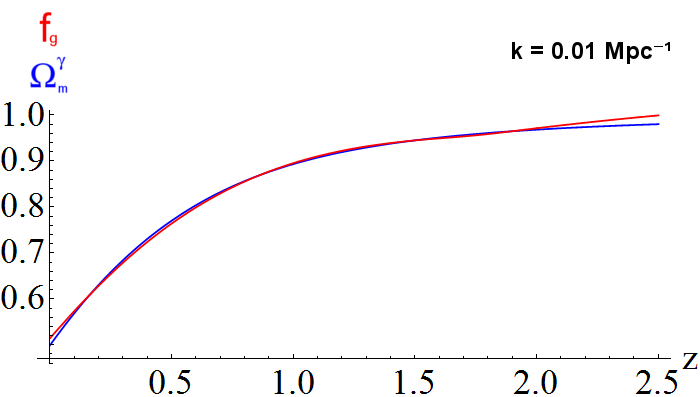}}
    \qquad
    \subfigure[\ model I]{\includegraphics[width=0.41\textwidth]{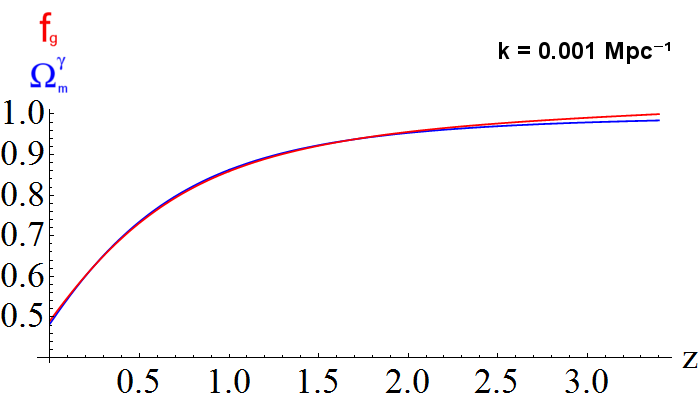}}
    \qquad
    \subfigure[\ model II]{\includegraphics[width=0.41\textwidth]{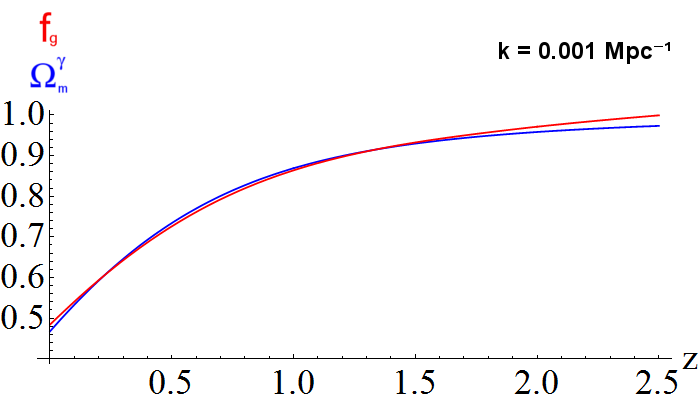}}
    \caption{Cosmological evolutions of the growth rate $f_\mathrm{g}$ (red line) and $\Omega_\mathrm{m}^\gamma$ (blue line) with $\gamma = \gamma_0
    + \gamma_1 \cdot \frac{z}{1 + z}$ as functions of the redshift $z$ in model I for $k = 0.1 \mathrm{Mpc}^{-1}$ (a), $k = 0.01 \mathrm{Mpc}^{-1}$
    (c) and $k = 0.001 \mathrm{Mpc}^{-1}$ (e), and those in model II for $k = 0.1 \mathrm{Mpc}^{-1}$ (b), $k = 0.01 \mathrm{Mpc}^{-1}$ (d) and
    $k = 0.001 \mathrm{Mpc}^{-1}$ (f).}
    \label{figure_rational_growth_index}
\end{figure}
        %%%%%%%%%%%%%%%%%%%
        %%%%%%%%%%%%%%%%%%%
        %%%%% Fig. 13 %%%%%
        %%%%%%%%%%%%%%%%%%%
\begin{figure}[!h]
    \subfigure[\ model I]{\includegraphics[width=0.45\textwidth]{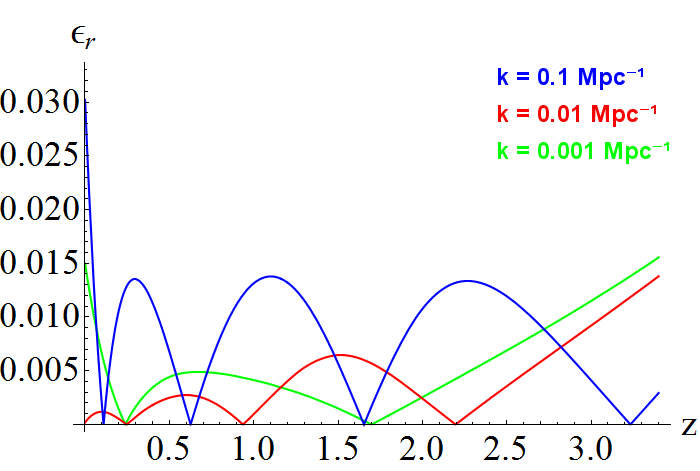}}
    \quad
    \subfigure[\ model II]{\includegraphics[width=0.45\textwidth]{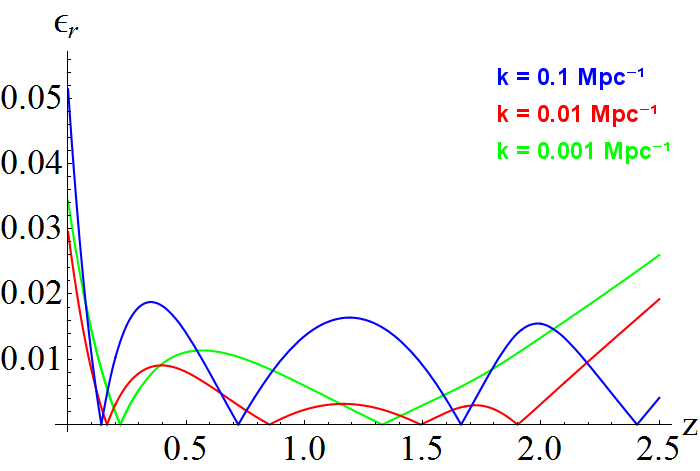}}
    \caption{Cosmological evolution of the relative difference $\epsilon_r = \frac{\left| f_\mathrm{g} - \Omega_\mathrm{m}^\gamma \right|}{f_\mathrm{g}}$ with
    $\gamma = \gamma_0 + \gamma_1 \cdot \frac{z}{1 + z}$ for $k = 0.1 \mathrm{Mpc}^{-1}$ (blue line), $k = 0.01 \mathrm{Mpc}^{-1}$ (red line) and $k = 0.001 \mathrm{Mpc}^{-1}$
    (green line) in model I (a) and model II (b).}
    \label{rat_rel_dif}
\end{figure}
        %%%%%%%%%%%%%%%%%%%

    The cosmological evolutions of the growth rate $f_\mathrm{g}(z)$ and $\Omega_\mathrm{m}(z)^{\gamma(z)}$ in model I and model II for several values of $k$ are
    depicted in Fig.~\ref{figure_rational_growth_index}. The fits in this case improve the results obtained for a constant growth index, but they seem similar to
    those obtained for the linear case.

    As in the previous subsections, the relative difference $\epsilon_r$ for several values of $k$ in model I and model II is shown in Fig.~\ref{rat_rel_dif} in
    order to analyze the fits quantitatively. For model I the relative difference is less than $3\%$ for $\log k(\mathrm{Mpc}^{-1}) = - 1$ and $1.5\%$ for
    $\log k(\mathrm{Mpc}^{-1}) \leq - 2$; while for model II, the highest value of $\epsilon_r$ is $5\%$ for $\log k(\mathrm{Mpc}^{-1}) = - 1$, $3\%$ for
    $\log k(\mathrm{Mpc}^{-1}) = - 2$ and $3.5\%$ for $\log k(\mathrm{Mpc}^{-1}) = - 3$. These results improve those obtained for the case given by
    $\gamma = \gamma_0$. For model I the results are very similar to the corresponding ones of $\gamma = \gamma_0 + \gamma_1 \cdot z$, but for model II they are
    worse.

    To conclude, it is important to point out that three parametrizations for the growth index have been studied for both model I and model II. As it may have been
    expected, the fits obtained for a constant growth index give the worst results. The results obtained for the other two ansatz considered, i.e. $\gamma = \gamma_0 +
    \gamma_1 \cdot z$ and $\gamma = \gamma_0 + \gamma_1 \cdot \frac{z}{1+z}$, are quite similar for model I, but $\gamma = \gamma_0 + \gamma_1 \cdot z$ gives better
    fits for model II than those corresponding to the ansatz $\gamma = \gamma_0 + \gamma_1 \cdot \frac{z}{1+z}$. In conclusion, the linear ansatz, $\gamma =
    \gamma_0 + \gamma_1 \cdot z$, is the best parametrization for the growth index for the two models considered.

%%%%%%%%%%%%%%%%%%%%%%%%%%%
%%%  Sec. VI
%%%%%%%%%%%%%%%%%%%%%%%%%%%
\section{Discussion}

    In this section, the content of the paper is summarized and the results obtained for model I and model II are analyzed.

    Two models of $F(R)$ modified gravity given by (\ref{iii2}) and (\ref{iii5}) have been considered throughout this work. The parameters of these models have
    been set in order to agree with current observational data coming from \cite{Komatsu:2010fb}. Model (\ref{iii2}) with the set of parameters given by
    (\ref{iii7}) is the so-called model I, while model (\ref{iii5}) with the set of parameters given by (\ref{iii9}) is the so-called model II.

    The growth of matter density perturbations has been studied for model I and model II. The so-called growth rate has been obtained numerically for both models
    and three ansatz for the so-called growth index have been considered. In Figs.~\ref{constant_growth_index_vs_logk}, \ref{linear_growth_index_vs_logk} and
    \ref{rational_growth_index_vs_logk}, the results of the different parametrizations for the growth index are shown for both models.

    To determine which ansatz of those considered for the growth index fits better Eq.~(\ref{iv5}) to the solution of Eq.~(\ref{iv4}), the results obtained
    for the three parametrizations have been analyzed in the previous section. The ansatz given by $\gamma = \gamma_0 + \gamma_1 \cdot z$ seems to be the best
    choice for both models.

\begin{figure}[!h]
    \subfigure[]{\includegraphics[width=0.45\textwidth]{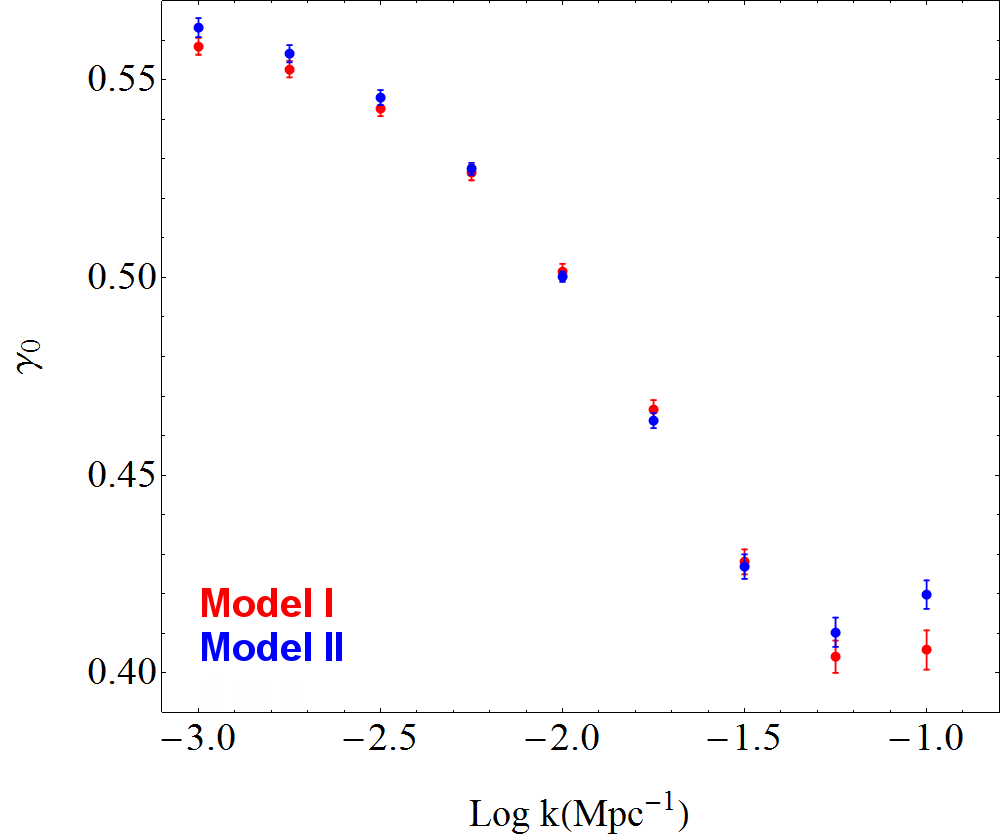}}
    \quad
    \subfigure[]{\includegraphics[width=0.45\textwidth]{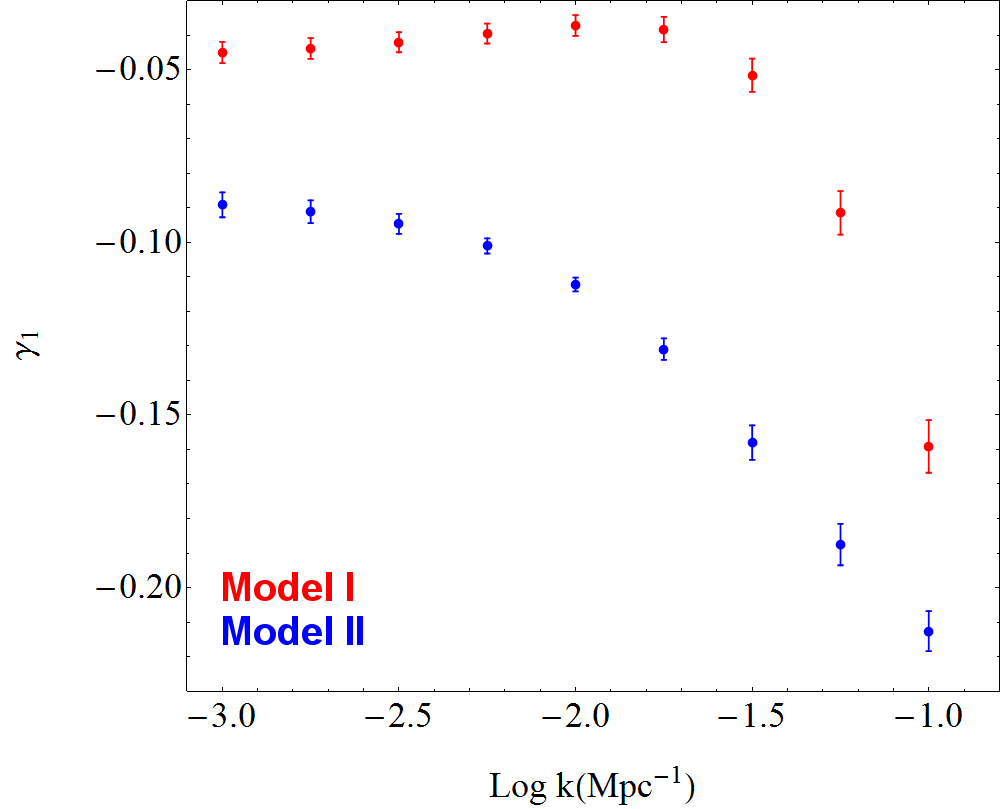}}
    \caption{Growth index fitting parameters in the case $\gamma = \gamma_0 + \gamma_1 \cdot z$ as a function of $\log k(\mathrm{Mpc}^{-1})$. The legend is the same as
    Fig.~\ref{constant_growth_index_vs_logk}.}
    \label{comparisonII}
\end{figure}

    Thus, in order to discriminate between model I and model II (or with the models considered in \cite{Bamba:2012qi}) using the growth history, the values of
    $\gamma_0$ and $\gamma_1$ in $\gamma = \gamma_0 + \gamma_1 \cdot z$ could have an essential importance. In Fig.~\ref{comparisonII} these two parameters,
    $\gamma_0$ and $\gamma_1$, are depicted for the two models together. We see that the behavior of $\gamma_0$ is very similar for both models. However, the values
    for $\gamma_1$ are totally different for model I to model II and they could be used to discriminate between these two models.

    One final note must be made. As it was pointed out in Eq.(\ref{iv1}), the evolution of matter density perturbations for $F(R)$ modified gravity theories
    depends on the comoving wave number $k$, which does not occur in the framework of general relativity. Throughout this work, this fact has been confirmed, in
    the first place with the results obtained for the growth rate $f_\mathrm{g}$, and finally, with those obtained for each of the three parametrizations considered for
    the growth index $\gamma$. Nevertheless, these parametrizations do not depend on the comoving wave number $k$, and it may be very interesting in the future to
    propose some scale-dependent ansatz for these $F(R)$ modified gravity theories.

%%%%%%%%%%%%%%%%%%%%%%%%
%%%  Acknowledgments
%%%%%%%%%%%%%%%%%%%%%%%%
\section*{Acknowledgments}

  I would like to thank the referee for comments and criticisms that led to the improvement of a first version. I would also like to thank Sergei Odintsov for
  suggesting the problem and for giving me some ideas to carry out this task. I would like to acknowledge a JAE fellowship from CSIC.

\end{document}